\documentclass[conference]{IEEEtran}
\IEEEoverridecommandlockouts

\usepackage{cite}
\usepackage{amsmath,amssymb,amsfonts}
\usepackage{algorithmic}
\usepackage{graphicx}
\graphicspath{{./figures/}}
\usepackage{caption}
\usepackage{subcaption}
\usepackage{textcomp}
\usepackage{xcolor}
\usepackage{todonotes}
\usepackage{tabularx,booktabs}
\def\BibTeX{{\rm B\kern-.05em{\sc i\kern-.025em b}\kern-.08em
    T\kern-.1667em\lower.7ex\hbox{E}\kern-.125emX}}
    
\DeclareMathOperator*{\argmin}{arg\,min}
    
\begin{document}

\title{Map-aided annotation for pole base detection}

\author{\IEEEauthorblockN{Benjamin Missaoui$^1$}
\and
\IEEEauthorblockN{Maxime Noizet$^1$}
\and
\IEEEauthorblockN{Philippe Xu$^1$}
\thanks{$^{1}$The authors are with the Universit\'e de technologie de Compi\`egne, CNRS, Heudiasyc, France.
{\tt\small name.surname@hds.utc.fr}}%
}

\maketitle

\begin{abstract}
For autonomous navigation, high definition maps are a widely used source of information.
Pole-like features encoded in HD maps such as traffic signs, traffic lights or streetlights can be used as landmarks for localization.
For this purpose, they first need to be detected by the vehicle using its embedded sensors.
While geometric models can be used to process 3D point clouds retrieved by lidar sensors, modern image-based approaches rely on deep neural network and therefore heavily depend on annotated training data.
In this paper, a 2D HD map is used to automatically annotate pole-like features in images.
In the absence of height information, the map features are represented as pole bases at the ground level.
We show how an additional lidar sensor can be used to filter out occluded features and refine the ground projection.
We also demonstrate how an object detector can be trained to detect a pole base.
To evaluate our methodology, it is first validated with data manually annotated from semantic segmentation and then compared to our own automatically generated annotated data recorded in the city of Compi\`egne, France.
\end{abstract}
\noindent\fbox{%
    \parbox{.97\columnwidth}{%
        Erratum: In the original version\cite{IV23}, an error occurred in the accuracy evaluation of the different models studied and the evaluation method applied on the detection results was not clearly defined. In this revision, we offer a rectification to this segment, presenting updated results, especially in terms of Mean Absolute Errors (MAE).
    }%
}

\section{Introduction}
\label{sec:intro}

High definition (HD) maps provide a rich prior knowledge to automated navigation systems.
Many different types of information are encoded in HD maps, often organized in layers.
The road topology layer helps decision-making and motion planning while the features layer can be used for localization.
For the latter, there are various kinds of HD maps.

One type of maps consists in low-level features such as point clouds or image-based key points/frames.
These types of maps are often constructed from a SLAM perspective and their use may be restricted to the use of sensors similar to the ones employed during the mapping phase.
Another constraint comes from the scalability of such maps as they are often heavy in terms of storage.
However, one main advantage is that the mapped features being essentially geometric, the map can be constructed with a low level of human intervention.

Another type of maps consists in so-called vector map, which encodes features at a higher semantic level.
The features are typically road infrastructure such as traffic signs, traffic lights, road markings or sometimes building footprints.
The geometry of these features are often vectorized in 2D with geometric primitives such as points and lines.
This enables these maps to be relatively light and can be easily scaled up at a city level.
Contrary to the previous case, these maps often require a higher amount of intervention from a human operator and could be harder to update automatically.

\begin{figure}[t!]
 \centering
 \includegraphics[width=\columnwidth]{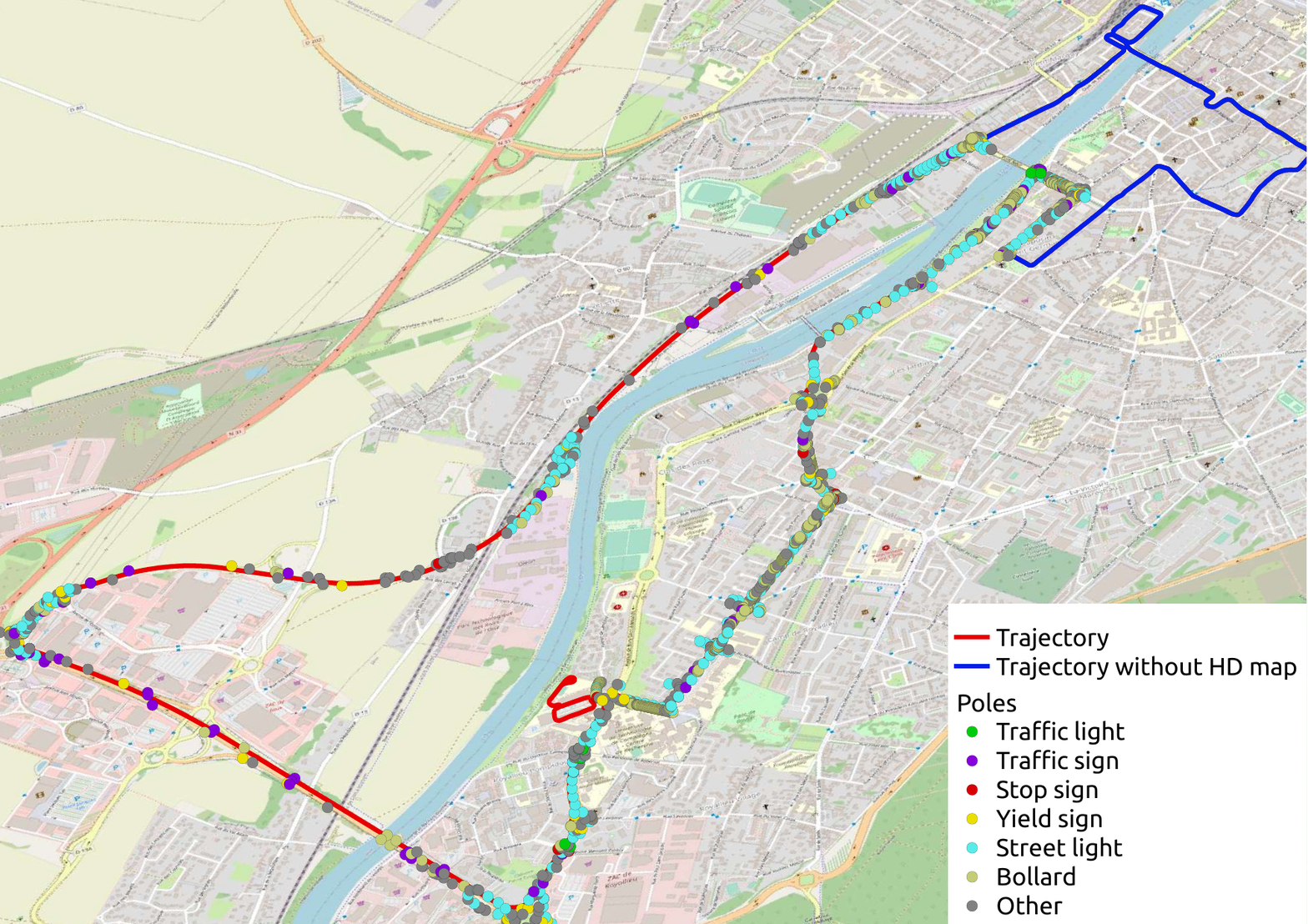}
 \caption{HD map of the city of Compi\`egne, France, containing generic pole-like elements such as traffic signs, bollards or street lights.}
 \label{fig:HD-map}
\end{figure}

In this work, we consider only vector maps as they are sensor agnostic and could be provided in generic frameworks such as OpenStreetMap.
The information encoded by road features such as traffic signs and traffic lights can be used for navigation but the knowledge of their spatial position can also be used for localization purposes.
More generally, all georeferenced features can serve as landmarks to estimate the vehicle pose.
To do so, the vehicle needs to be able to perceive these features from its own embedded sensors.
Fig.~\ref{fig:HD-map} shows an HD map of the city of Compi\`egne, France, containing several types of point features such as traffic signs or traffic lights among others.

Lidars and cameras are the most common sensors used in the field of intelligent vehicles.
Image-based methods for object detection or semantic segmentation have seen a tremendous improvement in past years thanks to deep neural networks.
A key element for these methods to work is to have a large amount of annotated data.

We propose in this paper a method to make use of HD maps for automatic annotation of images.
In particular, we show how the concept of pole base can be defined from road features and how they can be detected from an object detection point-of-view using state-of-the-art deep learning methods.
The contributions of this paper are as follows:
\begin{itemize}
    \item definition of a pole base class from HD map pole-like features
    \item projection, refinement and filtering of image annotations by the use of a lidar sensor
    \item bounding box annotations for object-based detection
    \item experimental validation using both manually annotated data from semantic segmentation datasets and from real data and an HD map in the city of Compi\`egne, France
\end{itemize}

The rest of the paper is organized as follows. 
First, some related works are presented in section~\ref{sec:sota}.
In section~\ref{sec:hdmap}, we introduce how HD map features can be projected onto images for automatic annotations and why additional refinement and filtering using a lidar are needed.
To compare with manually annotated data, we show in section~\ref{sec:segmentation} how semantic segmentation data can be used to achieve similar annotations.
Next, in section~\ref{sec:detector}, we introduce the use of a bounding-box based object detector for pointwise detection.
Finally, experimental results are detailed in section~\ref{sec:exp}, first using data from semantic segmentation datasets and then from real unannotated data along with an HD map acquired in the city of Compi\`egne.

\section{Related works}
\label{sec:sota}

The use of pole-like features in localization context is fairly common~\cite{li_robust_2021,sefati_improving_2017,spangenberg_pole-based_2016}.
The most common sensors used in intelligent vehicles context are cameras and lidars.
For lidars, geometric assumptions can be used in order to build pole-like object detectors~\cite{gouda_fully_2022,lehtomaki_detection_2010,rodriguez-cuenca_automatic_2015}.
The lidar point intensity can also be used to detect reflective traffic signs~\cite{ghallabi_lidar-based_2019}.
Contrary to lidars, modern camera-based detectors mostly rely on machine learning techniques especially through the use of deep neural networks.
Therefore, they heavily rely on available annotated data.
The detection of road features such as traffic signs or traffic lights can be made using object detection methods requiring bounding box annotations.
Many generic object detectors are trained from datasets such as the Microsoft COCO~\cite{lin14} or more specialized one such as KITTI~\cite{geiger12}.
The detection of pole-like features, however, such as street lights is often considered from semantic segmentation point-of-view.
This requires pixel-level annotations which are very costly.
Several large scale semantic segmentation datasets are nevertheless available such SemanticKITTI~\cite{Behley19}, Cityspaces~\cite{cordts16} or BDD100K~\cite{yu20}.

In order to be less reliant on manually annotated data, Dong et al.~\cite{dong23} proposed to use range images constructed from a lidar in order to detect pole-like features which then served as pseudo-labels to train a deep neural network.
Sun et al.~\cite{Sun20} used an HD map to generate image-level labels, e.g., number of lanes in an image.
Lee et al.~\cite{Lee21} proposed a semi-automatic traffic landmark annotation using 3D road features from an HD map.
Their goal was to provide initial annotations to accelerate a human annotation process.
In this paper, we aim at using 2D HD maps to automatically annotate pole-like features in images while using an object detection approach.

\section{HD map features projection and filtering}
\label{sec:hdmap}

Throughout this paper, the map features are considered to be encoded as georeferenced two-dimensional points without height information.
In order to apply geometric transformation to these point features, their geodetic coordinates, i.e., longitude and latitude, are first converted into an East-North-Up (ENU) coordinates frame with its origin being in the vicinity of the driving sequences.
Note that the ``up'' coordinate can be ignored as the altitude is set arbitrarily.
A map feature is therefore represented as a 2D point \({}^{\text{M}}P\) with coordinates \(\left[{}^{\text{M}}x, {}^{\text{M}}y\right]^\top\) expressed in the map frame denoted by the superscript \(\text{M}\).
The map is therefore a set of \(n\) 2D points \({}^{\text{M}}\mathcal{M}=\left\{{}^{\text{M}}P_i\right\}_{i=1,\ldots,n}\).

\subsection{Projection of pole base onto image frames}
In order to project the map features onto the image frame, they are first expressed in the vehicle frame which is defined as the body frame of the vehicle IMU placed above the center of the rear axle.
A map feature point \({}^{\text{V}}P\) in the vehicle frame is related to \({}^{\text{M}}P\) as follows
\begin{align}
 \begin{bmatrix}
  {}^{\text{V}}x \\ {}^{\text{V}}y
 \end{bmatrix} & =
 \begin{bmatrix}
  \cos\theta & \sin\theta \\
  -\sin\theta & \cos\theta
 \end{bmatrix}
 \begin{bmatrix}
  {}^{\text{M}}x-x \\ {}^{\text{M}}y-y
 \end{bmatrix},
 \label{eq:MtoV}
\end{align}
where \([x,y]^\top\) is the coordinates vector of the vehicle and \(\theta\) its heading.
Note that the accuracy of the coordinates of \({}^{\text{V}}P\) gets more sensitive with respect to the heading estimate \(\theta\) when the map features \({}^{\text{M}}P\) gets further away from the vehicle position.

Because there is no height information available, it is not possible to project the map features directly onto an image.
Instead, we decided to consider all the map features at ground level.
That is to say, regardless of the true height of the map feature, we will only consider its projection onto the ground.
For pole-like features such as bollards or street lights, it will correspond to their base, i.e., the contact point with the ground.
For traffic signs, most of them are actually fixed on top of poles although these poles may not be mapped explicitly.
Therefore, all the map features will be considered as represented in the image frame by their respective pole base point.

By assuming that the \(x\)-\(y\) plane of the vehicle frame is parallel to the ground and at fixed known height \(h\), a map feature can be set at the ground level by setting 
\begin{align}
    {}^{\text{V}}z:=-h.
    \label{eq:height}
\end{align}
These points can then be projected onto the camera image by making use of the camera intrinsic calibration parameters as well as its extrinsic ones with respect to the vehicle frame.

\begin{figure}[t!]
 \centering
 \includegraphics[width=\columnwidth]{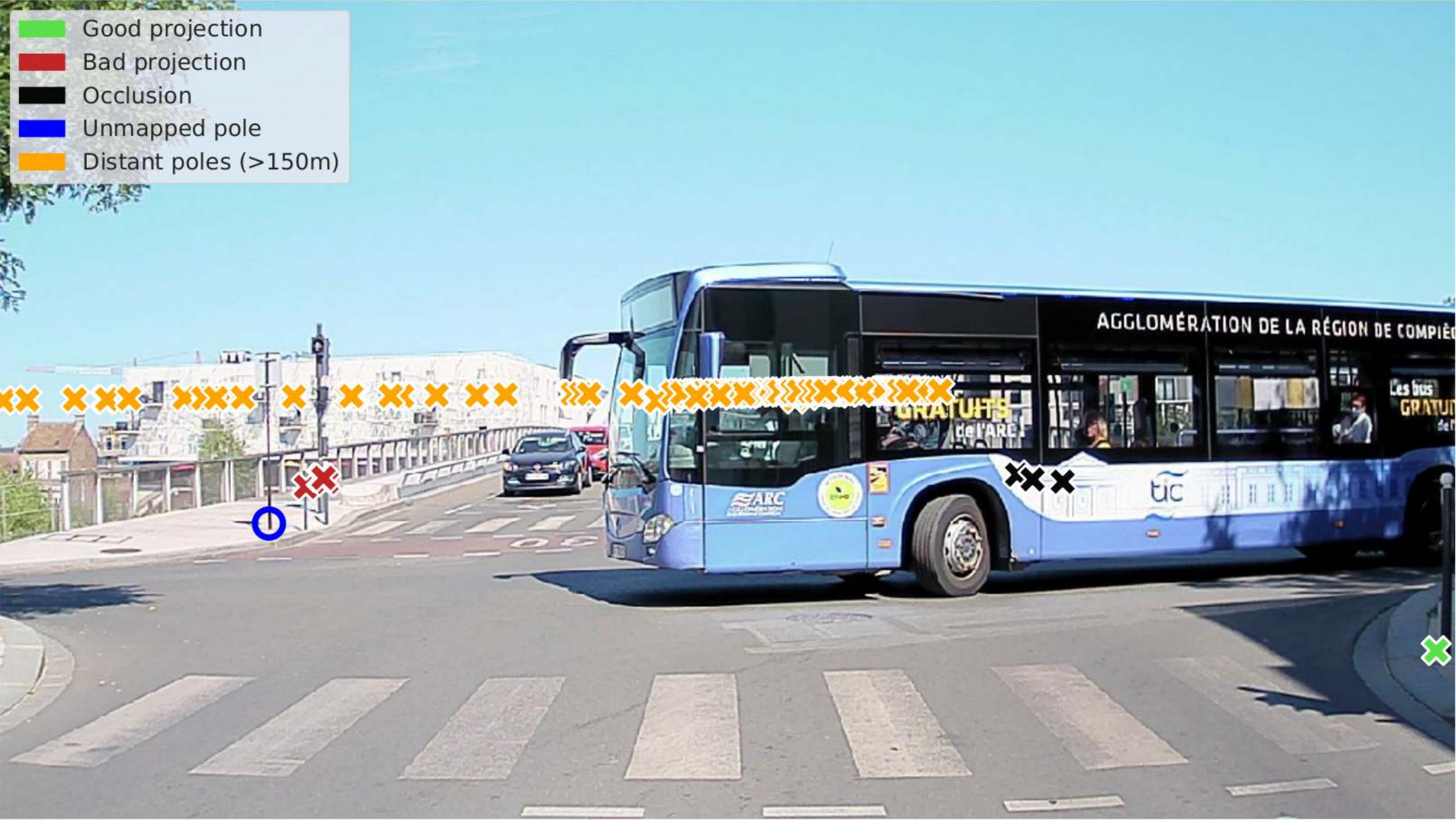}
 \caption{Naive projection of map features onto an image frame. Only the green point on the right-hand side can be considered as a correct annotation.}
 \label{fig:map-proj}
\end{figure}

Fig.~\ref{fig:map-proj} illustrates the projection of some map features onto an image.
The map feature points are supposed to represent the ground level projection of the map features.
The green point on the right side pictures is a well-projected map feature. 
The red ones represent ill-projected map features due to the parallel ground hypothesis being wrong.
The black points depict projected map features that are actually not visible from the camera due to occlusions.
The yellow points are distant features that are both occluded but also wrongly projected as they are all the more sensitive to the ground plane estimation.
Finally, the blue circle shows an example of an unmapped feature.
This shows that the projection of map features onto an image is not straightforward and needs further filtering and refinement.

\subsection{Lidar-based refinement and filtering}
In order to deal with the issues pictured in Fig.~\ref{fig:map-proj}, we propose to use an additional lidar sensor to refine and filter the map feature projection.
The lidar is used for two purposes.
The first is to better estimate the ground surface on which to project the map features and the second is to assess whether or not a feature is visible in the image frame.

For this purpose, we make use of a Hesai Pandora sensor, which provides a 3D lidar sensor placed on top of five cameras: four grayscale wide-angle cameras to cover a 360\({}^{\circ}\) field-of-view and an additional front color camera.
This setup allows to have almost the same field-of-view for both the lidar and the cameras.
That is to say that if a feature is occluded from the camera perspective, it is also the case from the lidar one.


Although the two modalities share the same field-of-view, the lidar sensor is limited in terms of range.
In our case, we consider a threshold of 150 meters above which we assume that the lidar cannot be used.
Therefore, a first step was to remove the map features outside a radius of 150 meters around the vehicle.
Note that these features could nevertheless still be visible in the image frame.
This first filtering allows to remove the yellow crosses pictured in Fig.~\ref{fig:map-proj}.

Next, the 3D point cloud \(\mathcal{C}=\left\{p_i=\left[ x_i, y_i, z_i \right]^\top\right\}_{i=1,\ldots,m}\) from the lidar is used to have a better estimate of the ground surface.
In this work, we use the Patchwork++ ground segmentation method proposed in~\cite{lee22}.
This algorithm separates all the lidar points into two groups, a ground points one \(\mathcal{G}\) and a non-ground one \(\overline{\mathcal{G}}\).
The group composed of ground points is used to estimate the height of the map features.
To do so, the map feature point \({}^{\text{V}}P\) in the vehicle frame is transformed into a point \({}^{\text{L}}P\) in the lidar frame using a similar equation as in \eqref{eq:MtoV} with the extrinsic calibration parameters encoding the relative position of the lidar w.r.t. the vehicle frame.
Similarly to the vehicle localization, the calibration quality of the lidar, especially in terms of angle, has a higher impact of map features far away from the vehicle.
Once a map feature has been projected in the lidar frame from a top down 2D space, we use a nearest neighbor approach to select the closest lidar point, categorized as belonging to the ground, and use its \(z\)-coordinate as a height estimate instead of the one from \eqref{eq:height}:
\begin{align}
    {}^{\text{L}}z := z_k, \text{ with } k = \argmin_{i,\ p_i\in\mathcal{G}} \left\Vert {}^{\text{L}}P - p_i\right\Vert_{2D},
    \label{eq:ground}
\end{align}
where the \(\left\Vert\cdot\right\Vert_{2D}\) operator is the Euclidean distance but using only the \(x\)-\(y\) coordinates.

\begin{figure}[t!]
 \centering
 \includegraphics[width=\columnwidth]{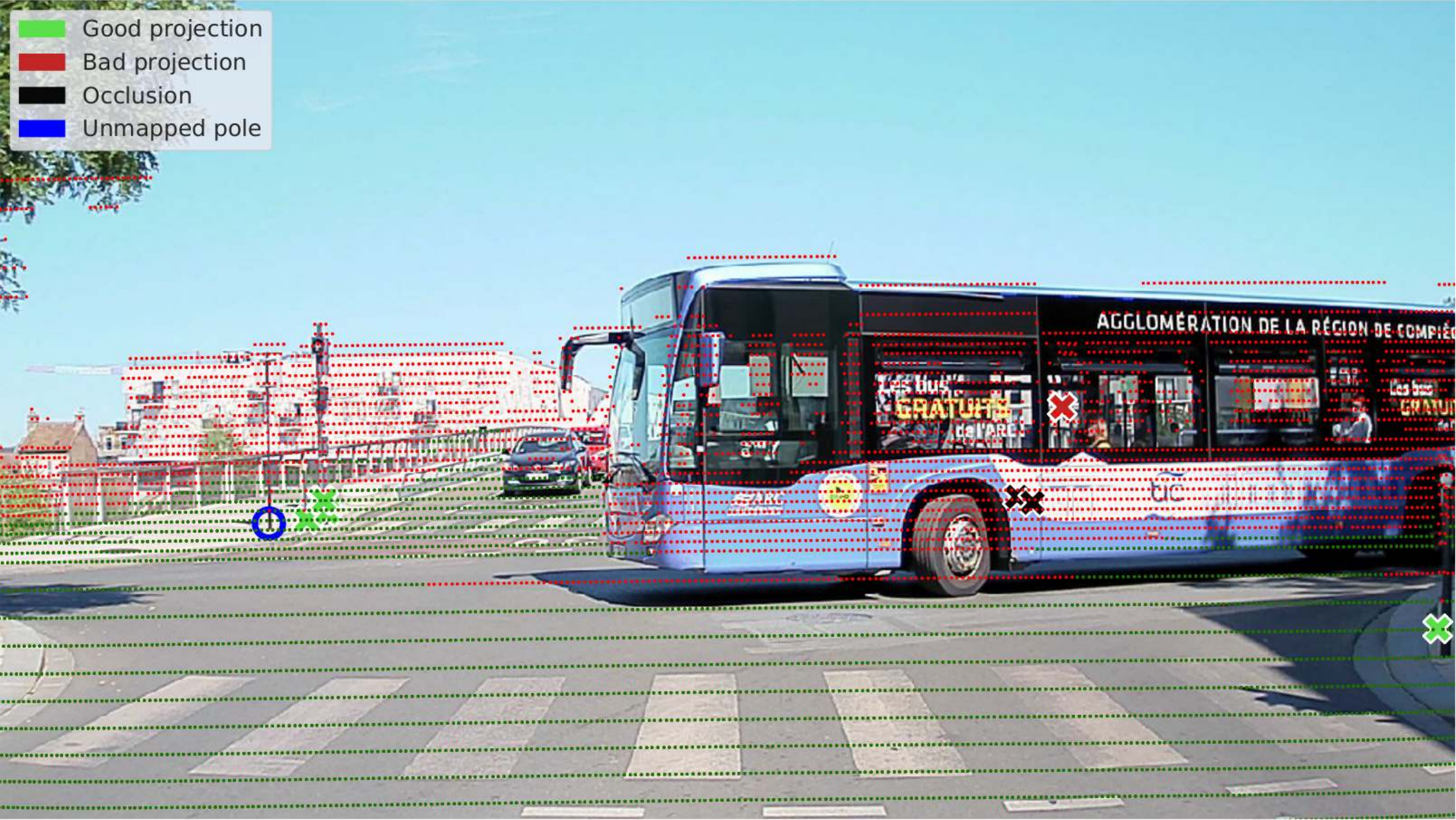}
 \caption{Projection of map features onto an image frame using Patchwork++.}
 \label{fig:map-proj-patchwork}
\end{figure}

Fig.~\ref{fig:map-proj-patchwork} shows the lidar point cloud projected onto the image.
The green dots correspond to point segmented as ground while the others are in red.
Using this new height estimation, the red crosses on the left side in Fig.~\ref{fig:map-proj} are now correctly projected at the ground level in Fig.~\ref{fig:map-proj-patchwork}.

The final step is to remove the map features that are occluded by some obstacles.
To do so, we first project all the lidar points onto the image as well as the map features.
For each map feature, we estimate the average depth using the lidar points that surround it in the image frame within a given radius.
This depth is then compared with the true distance separating the map feature and the camera frame.
The idea is that if a map feature is occluded by the obstacle, such as the bus in Fig.~\ref{fig:map-proj-patchwork}, then the estimated depth will be shorter than the true distance.
By setting a threshold on the distance difference, the occluded features can be removed.
The black crosses in Fig.~\ref{fig:map-proj-patchwork} are correctly classified as occluded and can be filtered.

These different steps enable to have a cleaner projection of the map features.
However, we can see in Fig.~\ref{fig:map-proj-patchwork} that some errors still remain.
Because the ground segmentation can also be erroneous some wrong projection can still occur.
We can also see the projection of the lidar onto the image is not perfect despite using the calibration parameters provided by the sensor manufacturer.
And finally, the unmapped features remain unannotated.

\section{Point annotation from semantic segmentation}
\label{sec:segmentation}

The automatic annotation presented in the previous section is still imperfect due to various noises from localization, calibration, map errors or imperfect ground estimation.
In order to assess the performance of a deep learning method for the detection of pole base features, we also study the use of manually labeled data.

To our knowledge, there exist no dataset with such annotations.
Fortunately, pole-like features can be found in datasets dedicated to semantic segmentation.
In this study, we consider the BDD100K dataset~\cite{yu20}.
In this dataset, we merge the three categories ``pole'', ``traffic sign'' and ``traffic light'' to represent our map features.
The annotations are provided at the pixel level, while in our case, we wish to mimic the point-wise annotations retrieved from an HD map.
Therefore, for each cluster of annotated map feature pixels, we need to compute the position of the pole base.

\begin{figure}[t!]
 \centering
 \begin{subfigure}[b]{0.23\textwidth}
  \includegraphics[width=\columnwidth]{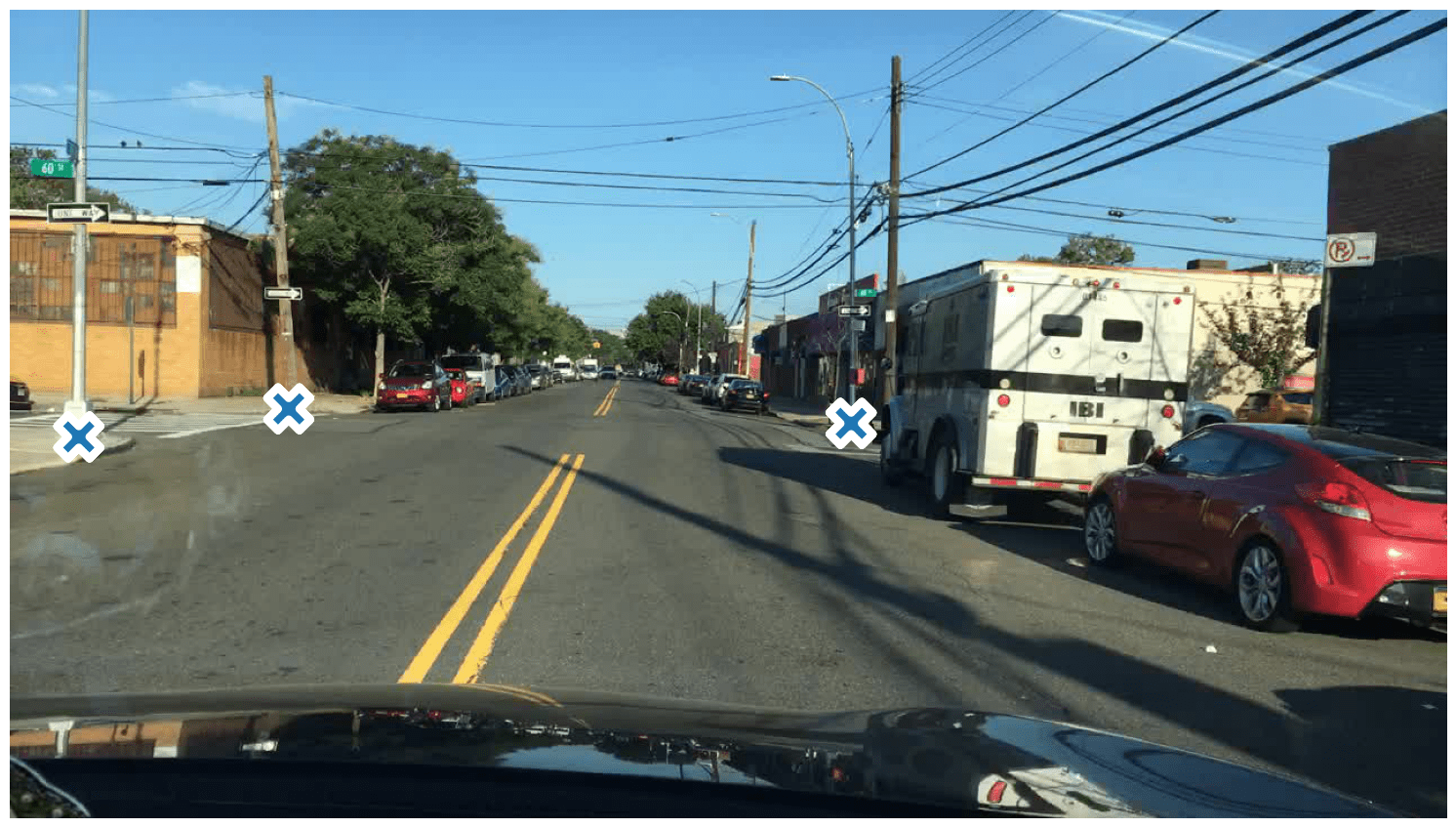}
  \caption{Point annotation}
 \end{subfigure}
 \begin{subfigure}[b]{0.23\textwidth}
  \includegraphics[width=\columnwidth]{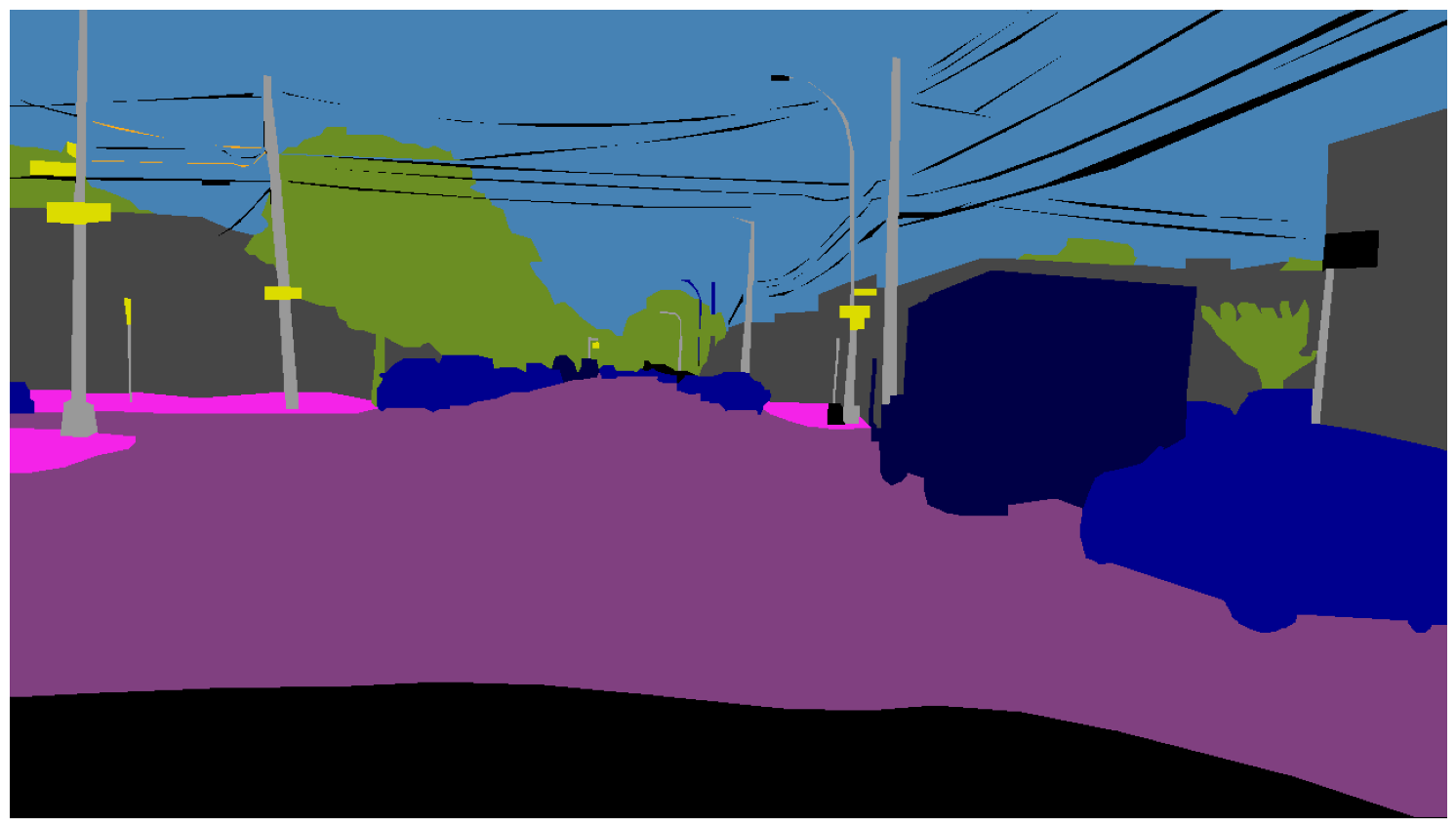}
  \caption{Semantic segmentation}
 \end{subfigure}
 \caption{Point annotations from semantic segmentation.}
 \label{fig:semantic}
\end{figure}

Like in the previous section, the base of a pole may not be visible due to occlusions.
Thanks to the pixel-wise annotations, we can define a pole base as the lower part of a map feature cluster that lies on top of a ground pixels.
The ground is defined by merging the following classes: ``road'', ``sidewalk'' and ``terrain''.
In addition to filtering out occluded poles, we also added a minimal width to the pixel clusters so as to filter out far away features.
Fig.~\ref{fig:semantic} illustrates some results on the BDD100K datasets.
The crosses represent the pole bases lying on the ground.

\section{Pole base detector}
\label{sec:detector}

In this study, we propose to formalize the detection of the pole base as an object detection problem.
Therefore, bounding boxes are used to represent the pole base.
At first glance, using a box to encode a point may not be relevant, especially for pole-like features that have thin appearances.
In our case, the boxes are not ``bounding'' an object as in classical object detection approaches.
Here, the boxes are used to encode the context surrounding the pole base while the centers of the boxes represent the points we aim at detecting.
The size of the boxes is a meta-parameter studied in the experimental process.
Using a bounding box formalism also allows to make use of highly efficient object detectors such as YOLO.
In this paper, the YOLOv7~\cite{wang22} algorithm is used.

To evaluate the performance of the detector, we consider two types of metrics.
First, the traditional object detection metrics are used (mAP, precision and recall), in order to evaluate the ability of the model to predict the right boxes. Note that we will be using the notation mAP 0.5:0.95 throughout this article, which refers to the mAP at different IoU thresholds, a metric commonly used in object detection as less saturated. Then, to have a finer evaluation of the point-wise accuracy of the box centers, we compute the horizontal distance with respect to the point annotations.
This choice is motivated by the fact that within a monocular-camera localization context, the horizontal coordinate of the detection points in the image frame can be used for bearing-only localization.
Therefore, the accuracy along the horizontal axis becomes more important than along the vertical one.
Nevertheless, a Euclidean 2D distance could also be used instead to compute the point-wise accuracy.

\section{Experimental results}
\label{sec:exp}

Our experiments were divided into two parts.
First, we validated the YOLO-based pole base detector using data from the BDD100K dataset.
These two cases represented the ideal context where the pole base features could be assumed to be perfectly annotated manually.
Then in a second step, we used an HD map to automatically annotate images.
This was done using an experimental Renault ZOE car equipped with a Pandora sensor and driving in the city of Compi\`egne, France.

\subsection{BDD100K}
Before trying to learn a pole base detector using our HD map aided automatic annotations, that is known to be partially imperfect, we first validated it using data from the BDD100K dataset. 
The annotation procedure described in Sec.~\ref{sec:segmentation} was applied to all the images from each dataset.
In addition, for the validation set, the annotations were manually reviewed in order to remove some remaining erroneous annotations generated from the semantic segmentation ground truth.
Table \ref{tab:bdd100knumbers} gives the number of training and validation examples for each dataset. 

\newcolumntype{Y}{>{\centering\arraybackslash}X}
\begin{table}[t!]
\caption{\label{tab:bdd100knumbers}BDD100K dataset information} 
\smallskip
\begin{tabularx}{0.48\textwidth}{l*{5}{Y}}
 \toprule     
         & Number of images & Number of poles\\
 \cmidrule(lr){2-3}
 Training      & 7628             & 9357              \\
 Validation & 372              & 909             \\
\bottomrule
\end{tabularx}
\end{table}




As stated in the previous section, we used an object detection approach with bounding boxes.
The pole base features were encoded by a box centered on it and with fixed size.
The choice of the box width and height then arose.
We decided to  experiment on BDD100K with squared boxes of various sizes ranging from 60 to 400 pixels.
Employing the official implementation of YOLOv7 with default hyperparameters, we set the IoU threshold to 0.5 during the non-maximum suppression step for evaluation purposes.  The model was initialized with the weights from MS COCO. 
The training took 18h roughly on a single Tesla V100 32G GPU for 300 epochs.
Most of the training converged after 100 epochs.

The predictions are accompanied by a confidence score. The lower this score, the less likely it is that an object is actually present at the location predicted by the bounding box. We can reject predictions with confidence score below a given value to remove potential wrong detections and improve the precision at the potential expense of recall.

We chose arbitrarily to apply a confidence threshold of 0.25 as it is unlikely that these low-confidence predictions genuinely correspond to objects. Besides, in our specific application context, maintaining a satisfactory precision is crucial.

The detection metrics obtained with this threshold for all the models are detailed in Table~\ref{tab:bdd100k-detect-metrics}.
When the boxes were too small, typically less than 100 pixels, the model exhibited poorer performance than other models, especially in terms of mAP.
The performance of the detector gets better and stable starting from 100 px.

One plausible explanation for this could be that small boxes do not contain enough local context. 
When gradually increasing their size, the boxes aggregate more and more information about the surroundings of the pole until these bits of information become enough for the model to recognize and discriminate the pole.
In the context of BDD100K, the option to increase box size emerges as a viable strategy to enhance performance, particularly since poles exhibit sufficient spacing, mitigating the risk of potential missdetections. When the box size exceeds 100px, it appears to contribute to improved precision and mAP at the expense of diminished recall. However, the detection performance can vary a lot between two box sizes.

We studied the localization accuracy of the coordinates of the center of the boxes w.r.t. the pole base annotations.
As stated previously, we only focused on the horizontal accuracy along the \(x\)-coordinate in the image frame.
Using all the true positive predicted bounding boxes, we computed the Mean Absolute Error (MAE) between the predicted \(x\)-coordinate and the labeled one.
As visible in Table~\ref{tab:bdd100k-detect-metrics}, enlarging the box size introduces more uncertainty in the pole base position. 
Significantly increasing box size can enhance overall performance, as shown by the 400x400 model ranking as the best in mean Average Precision (mAP). But, it is imperative to note a substantial increase of MAE and this suggests a trade-off.

\newcolumntype{Y}{>{\centering\arraybackslash}X}
\begin{table}[t!]
\caption{\label{tab:bdd100k-detect-metrics}Detection metrics and MAE after 100 epochs of training with different box sizes on the validation set of BDD100K. Precision and recall obtained are given using the confidence threshold 0.25.} 
\smallskip
\begin{tabularx}{0.48\textwidth}{l*{5}{Y}}
 \toprule     
 Box size & mAP 0.5:0.95 &       Precision    & Recall & MAE (px)\\
 \cmidrule(lr){2-5}
 60x60           &  42.5 & 71.8\%          & 58.7\% & \textbf{2.28}\\
 80x80             & 44.8 & 74.2\%          & 55.0\% & \textit{2.87}\\
100x100            & 54.7 & 67.6\%          & \textbf{68.1}\% & 3.18\\
150x150            & 57.6 & \textit{76.5}\%          & 61.8\% & 4.23\\
200x200           & 58.2 & \textbf{76.7}\%         & 61.9\% & 5.29\\
250x250          & \textit{58.5} & 66.1\%          & \textit{65.6}\% & 7.90\\
300x300           & 55.6 & 71.9\%          & 63.7\% & 9.73\\
350x350            & 55.2  & 67.7\%          & 61.2\% & 12.5\\
400x400   & \textbf{60.4} & 73.2\% & 62.8\%& 15.8\\
\bottomrule
\end{tabularx}
\end{table}

In this study, we chose to focus on a detector with high precision in detection, prioritizing a low false positive rate, even at the cost of accepting a reasonable reduction in recall and localization accuracy. We consequently opted for the 200x200 model as it achieved the highest precision and ranked third in terms of mAP 0.5:0.95. It is important to note that another confidence threshold could have led to another decision.

\begin{figure}[t!]
 \centering
 \includegraphics[width=\columnwidth]{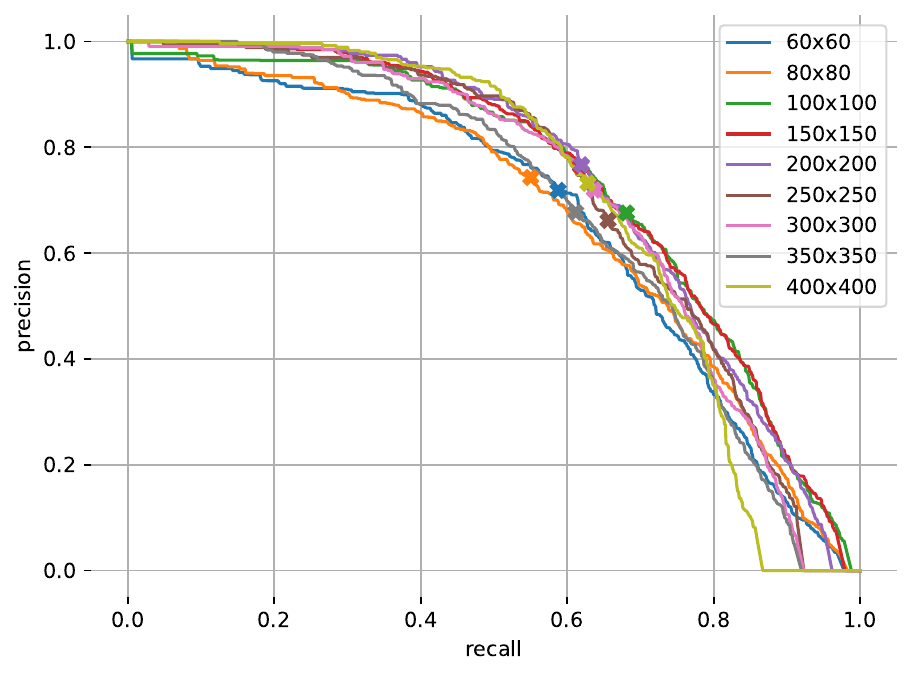}
 \caption{Precision-Recall curves obtained after 100 epochs of training with different box sizes on the validation set of BDD100K. The points corresponding to values obtained in Table~\ref{tab:bdd100k-detect-metrics} using a confidence threshold of 0.25 are visible using crosses.}
 \label{fig:precision-recall}
\end{figure}

Precision-Recall curves of all models are visible in Figure~\ref{fig:precision-recall}. It shows that the choice of the bounding box size depends in reality deeply on the precision-recall trade-off we want to apply.
The curves confirm that 60x60 and 80x80 on BDD100K seems insufficient to obtain sufficient recall when aiming a high precision. A minimal size of 100x100 is better for any trade-off and for a small increase of uncertainty.
Besides due to high MAE, 350x350 and 400x400 box sizes should be avoided, leading to a potential choice between 100x100 and 300x300 depending on the chosen trade-off.

As pictured in the figure, the arbitrary threshold applied previously corresponds to different precision-recall trade-offs depending on the model used. That is why, in the best scenario a precision or recall objective should be defined, then a box size and the corresponding threshold should be chosen to maximize the other metric while keeping in mind the localization accuracy objective.
The box size consequently become a new hyperparameter to reach the best detection performance.

\subsection{City of Compi\`egne}

\subsubsection{Data collection}
\begin{figure}[t!]
 \centering
 \includegraphics[width=\columnwidth]{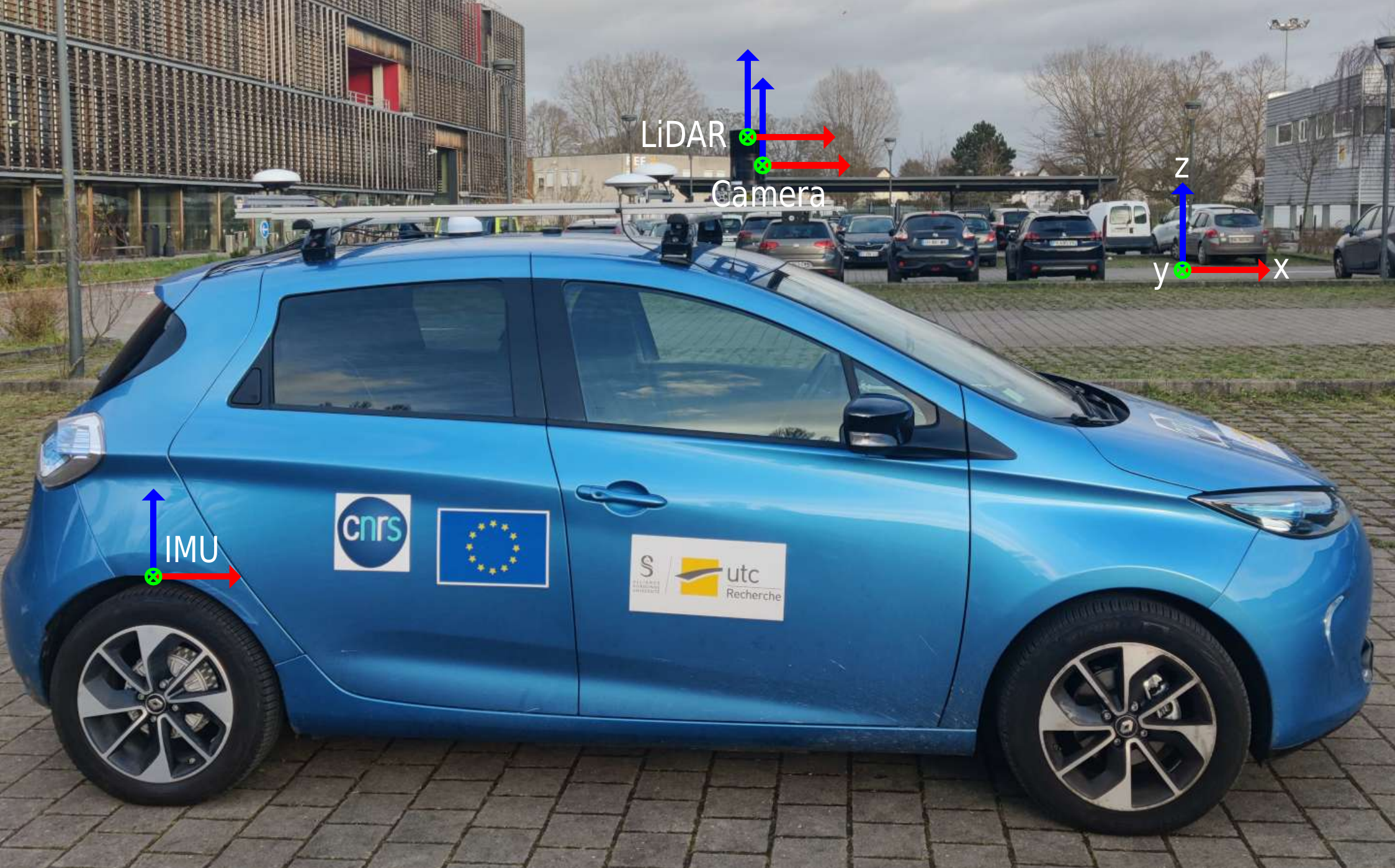}
 \caption{Experimental Renault ZOE vehicle equipped with a NovAtel SPAN-CPT GNSS/IMU and a Hesai Pandora sensor.}
 \label{fig:zoe-frame}
\end{figure}

An experimental Renault ZOE vehicle was equipped with the following sensors:
\begin{itemize}
    \item NovAtel SPAN-CPT GNSS/IMU with post-processed PPK computations for centimeter-level accuracy localization (50 Hz). 
    \item Hesai Pandora 40-layer LiDAR with a horizontal angular resolution of $0.2^\circ$ and integrating five synchronized cameras (10 Hz) (a front color camera, and four wide-angle mono cameras). For this work, only the color camera was used.
\end{itemize}
The extrinsic calibration between the Pandora lidar and the SPAN was obtained using a high-accuracy FARO Vantage laser tracker. For intrinsic calibration of cameras and the lidar, factory settings were used. 
Fig.~\ref{fig:zoe-frame} shows the experimental platform along with the different working frames.

The recorded sequence illustrated in Fig.~\ref{fig:HD-map} covered the entire city of Compi\`egne, France, with diverse road contexts. 
It lasted 40 minutes for a cumulative distance of 13.5 km.
This dataset was recorded within the context of the ERASMO~\cite{ERASMO23} European project.
The HD map used in this experiment was acquired in 2019.
In the meantime, some new poles were added and others were removed.
Additionally to annotation errors due to localization, calibration or projection, some other errors were also due to erroneous HD map data.

\subsubsection{Training sets}
In the driving sequence illustrated in Fig.~\ref{fig:HD-map}, the blue part of the trajectory was driven in an area where the environment was not mapped.
Therefore, we used this part of the sequence as the validation set and manually annotated the images in order not to bias the evaluation part.
For this part, only one image per second was kept and the images were further subsampled avoid having too many similar examples.
The rest of the sequence was used as our training set and the HD map was used to automatically annotate the images.

The sets obtained are described in Table \ref{tab:hdsnumbers}.
It is important to note that although the number of images seems larger compared to BDD100K, there is a lack of variability as they correspond to a unique driving sequence.
Similarly, the 6724 annotated poles actually correspond to a subset of 1600 unique poles in the HD map. 


\newcolumntype{Y}{>{\centering\arraybackslash}X}
\begin{table}[t!]
\caption{\label{tab:hdsnumbers}Heudiasyc dataset information} 
\smallskip
\begin{tabularx}{0.48\textwidth}{l*{5}{Y}}
 \toprule     
         & Number of images & Number of poles\\
 \cmidrule(lr){2-3}
 Train      & 15724             & 6724              \\
 Validation & 605              & 2148             \\
\bottomrule
\end{tabularx}
\end{table}

\subsubsection{Map-aided annotation}

To annotate our images using our HD map and the lidar, we set the parameters of the filtering described in Sec.~\ref{sec:hdmap} to the values detailed in Table \ref{tab:annotation_params}.

\newcolumntype{Y}{>{\centering\arraybackslash}X}
\begin{table}[t!]
\caption{\label{tab:annotation_params}Map-aided annotation parameters} 
\smallskip
\begin{tabularx}{0.48\textwidth}{l*{5}{Y}}
\toprule
 Maximum distance between a pole and the vehicle     & 150m\\
 Image search radius for pole lidar-based filtering & 20px\\
 Depth difference for pole lidar-based filtering & 5m\\
\bottomrule
\end{tabularx}
\end{table}

\begin{figure*}[t!]
 \centering
 \begin{subfigure}[b]{0.19\textwidth}
  \includegraphics[width=\columnwidth]{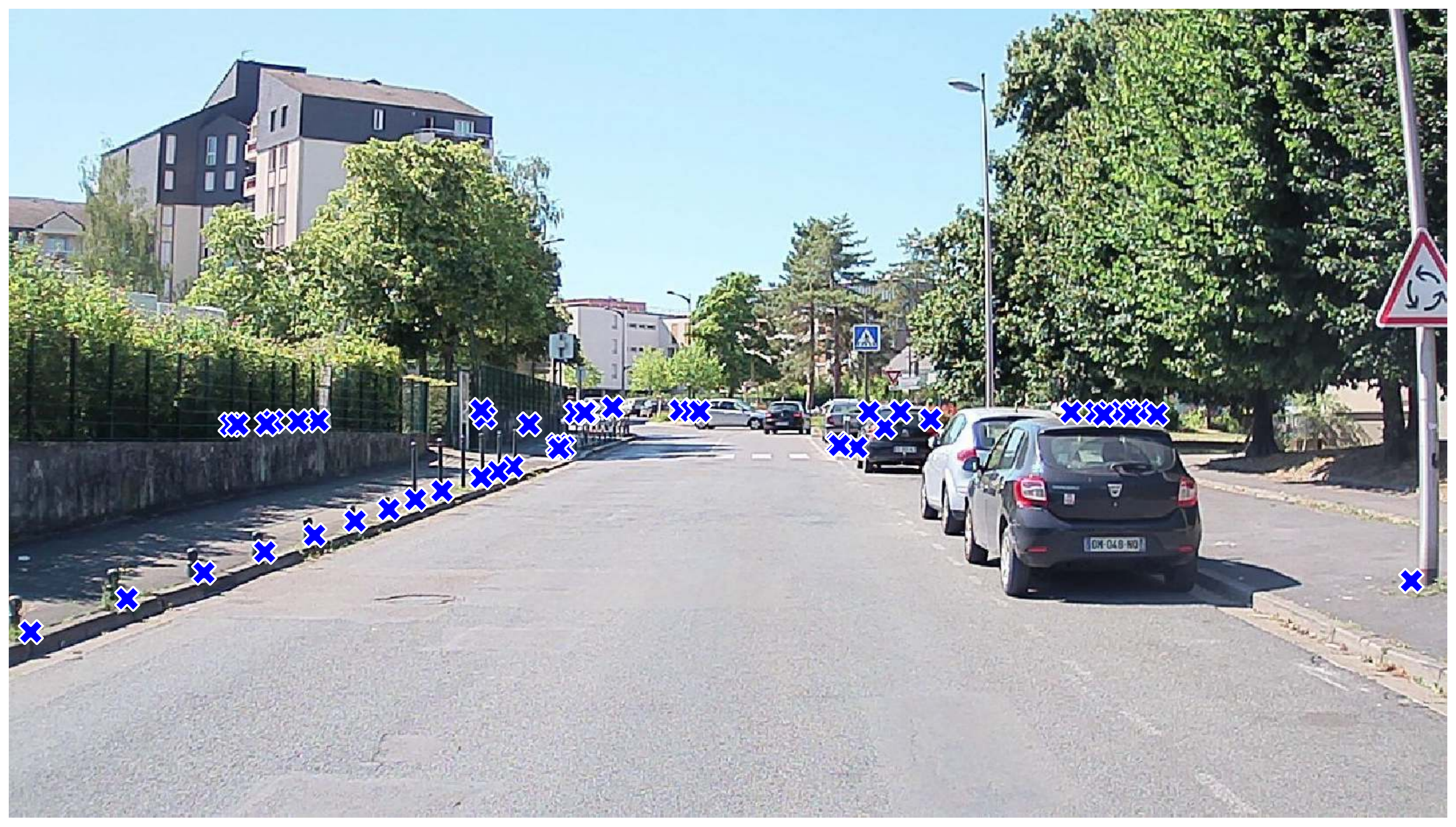}
 \end{subfigure}
 \begin{subfigure}[b]{0.19\textwidth}
  \includegraphics[width=\columnwidth]{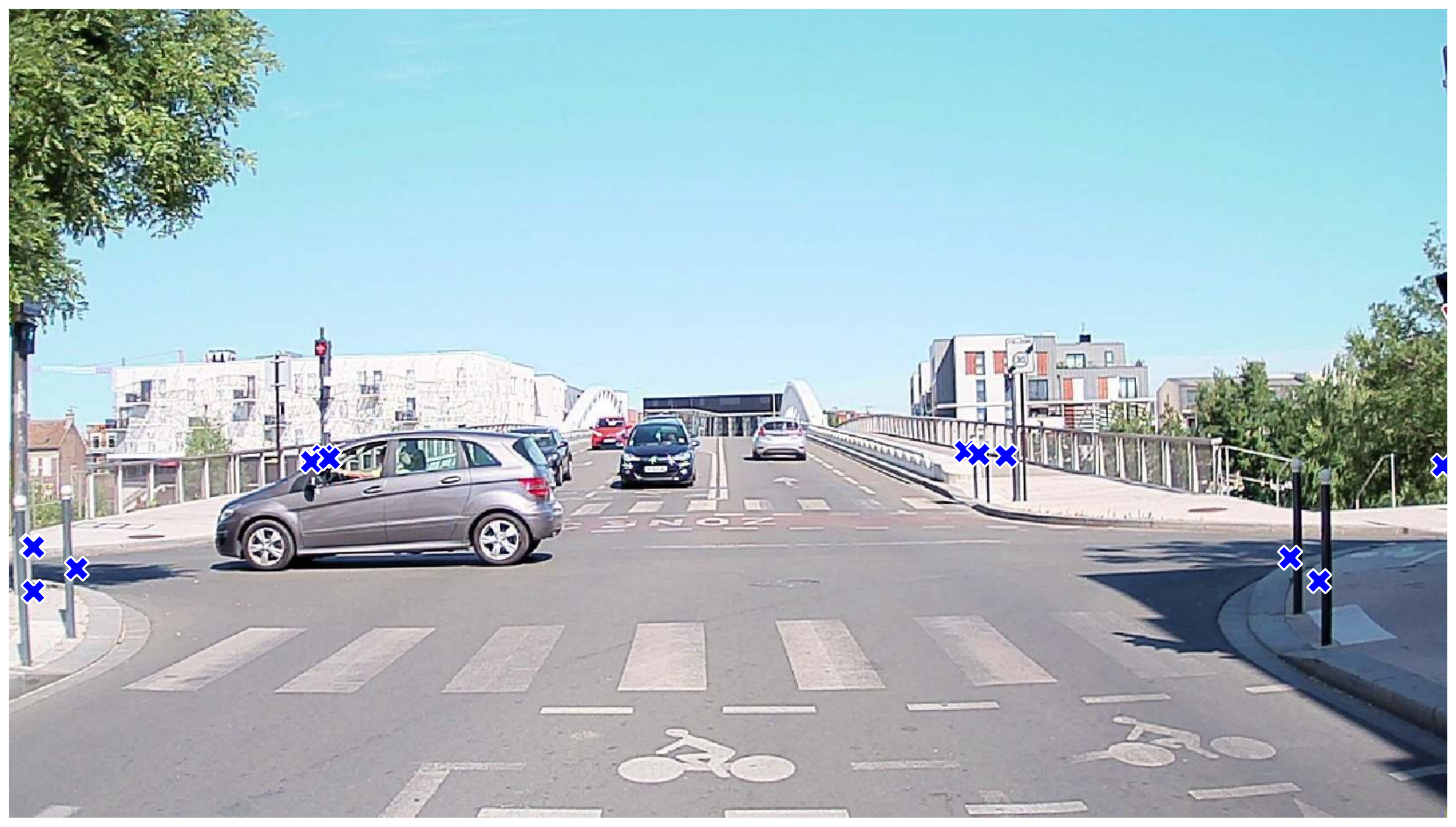}
 \end{subfigure}
 \begin{subfigure}[b]{0.19\textwidth}
  \includegraphics[width=\columnwidth]{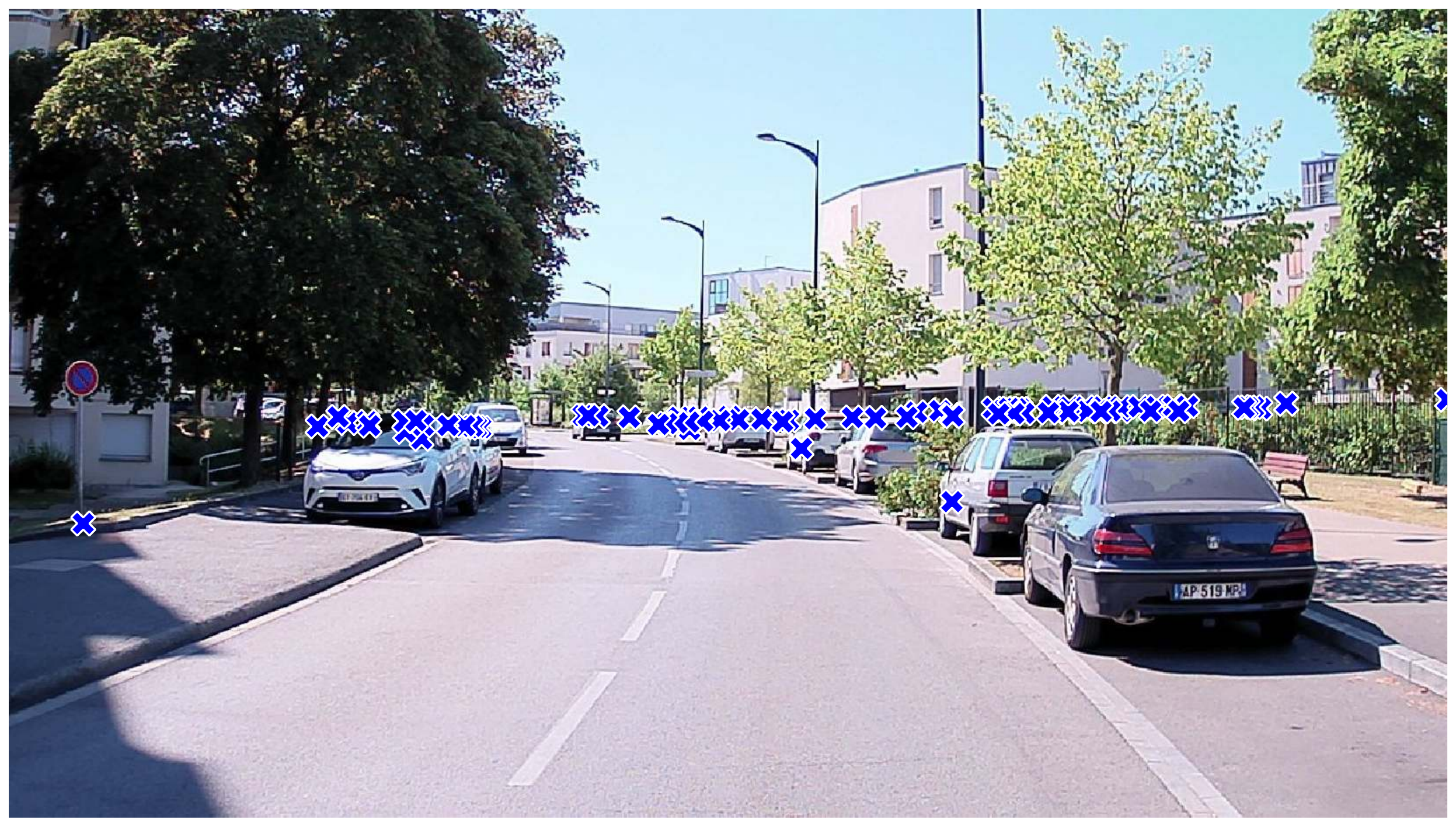}
 \end{subfigure}
 \begin{subfigure}[b]{0.19\textwidth}
  \includegraphics[width=\columnwidth]{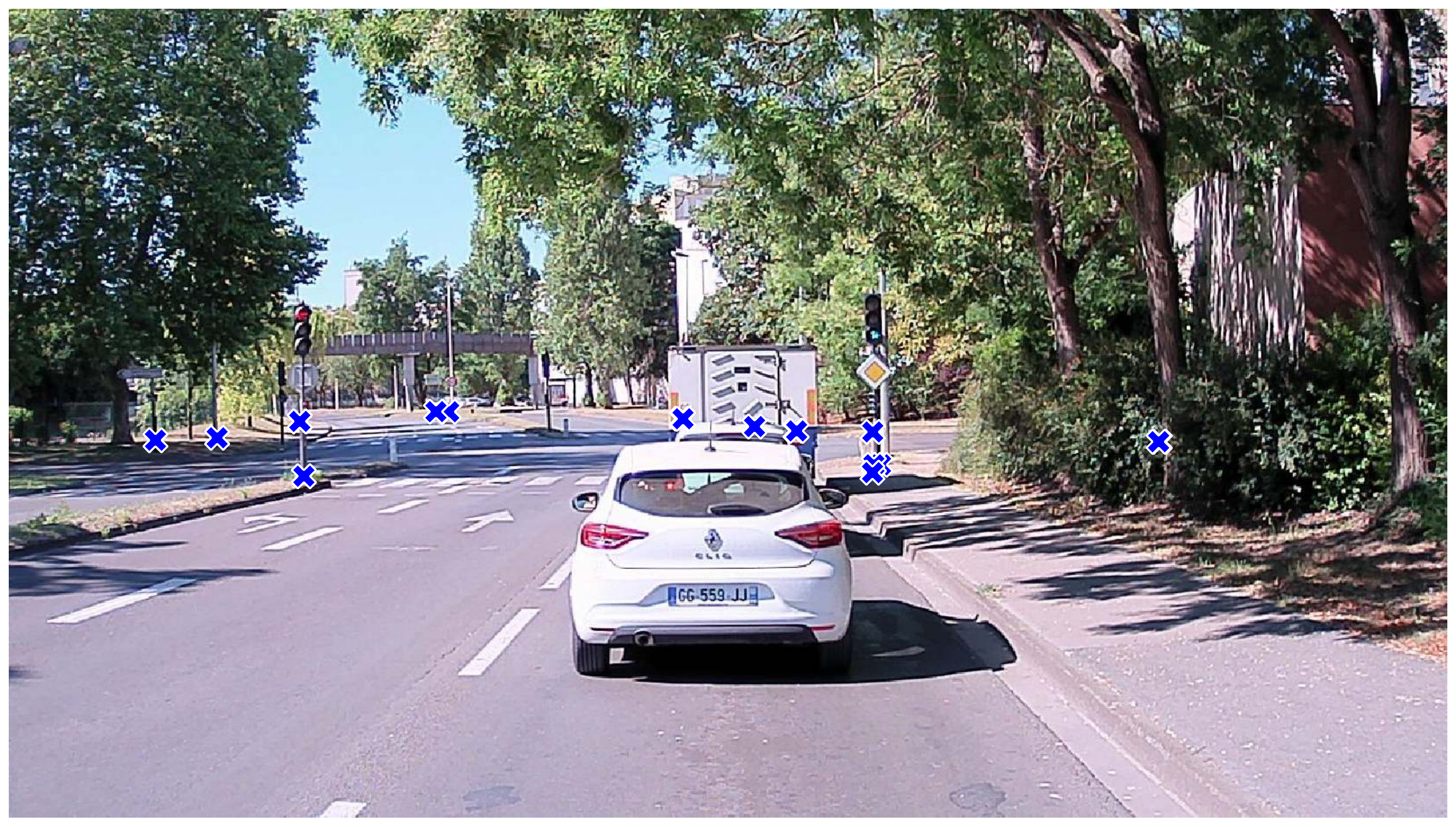}
 \end{subfigure}
 \begin{subfigure}[b]{0.19\textwidth}
  \includegraphics[width=\columnwidth]{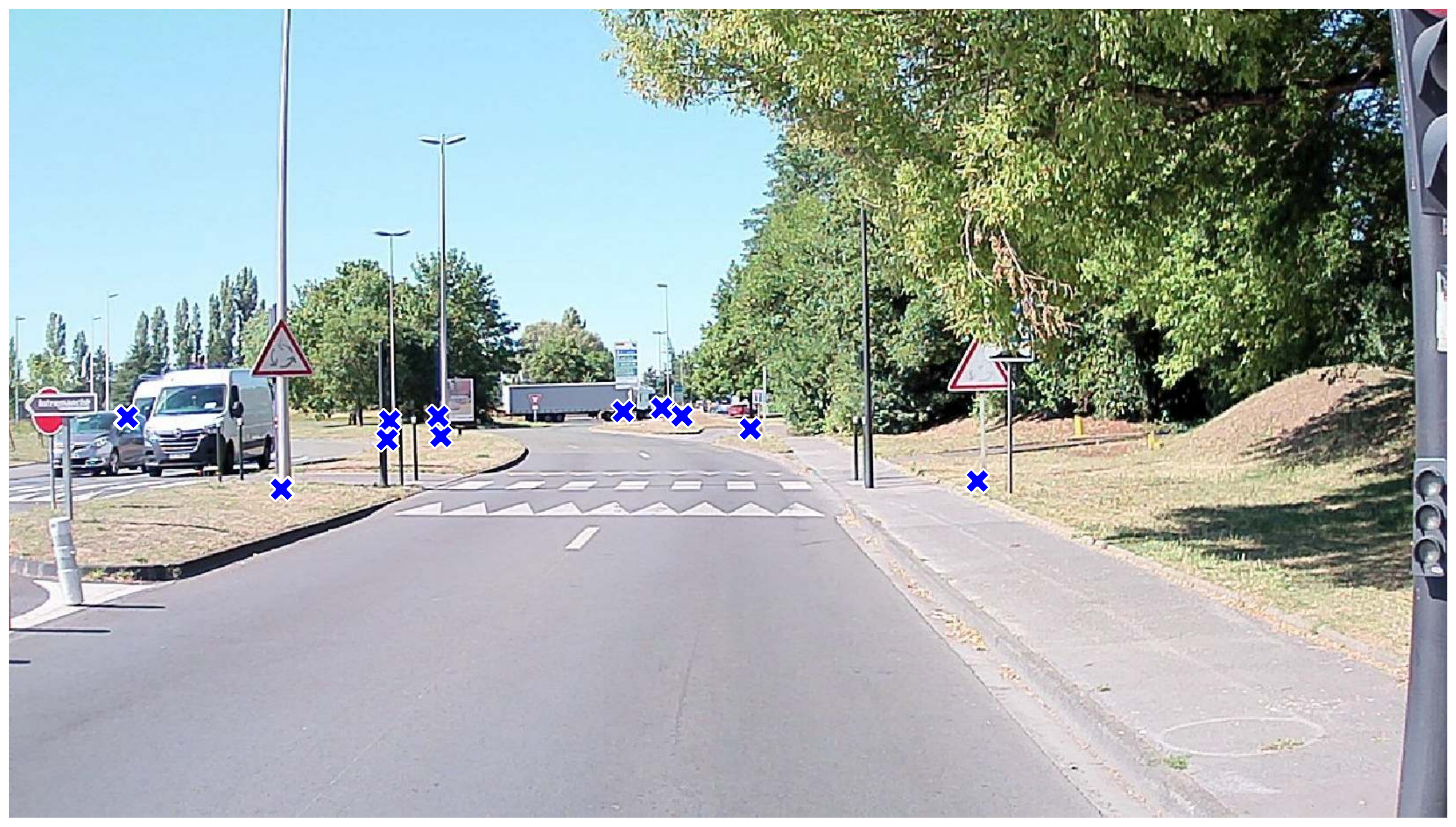}
 \end{subfigure}
 
 \begin{subfigure}[b]{0.19\textwidth}
  \includegraphics[width=\columnwidth]{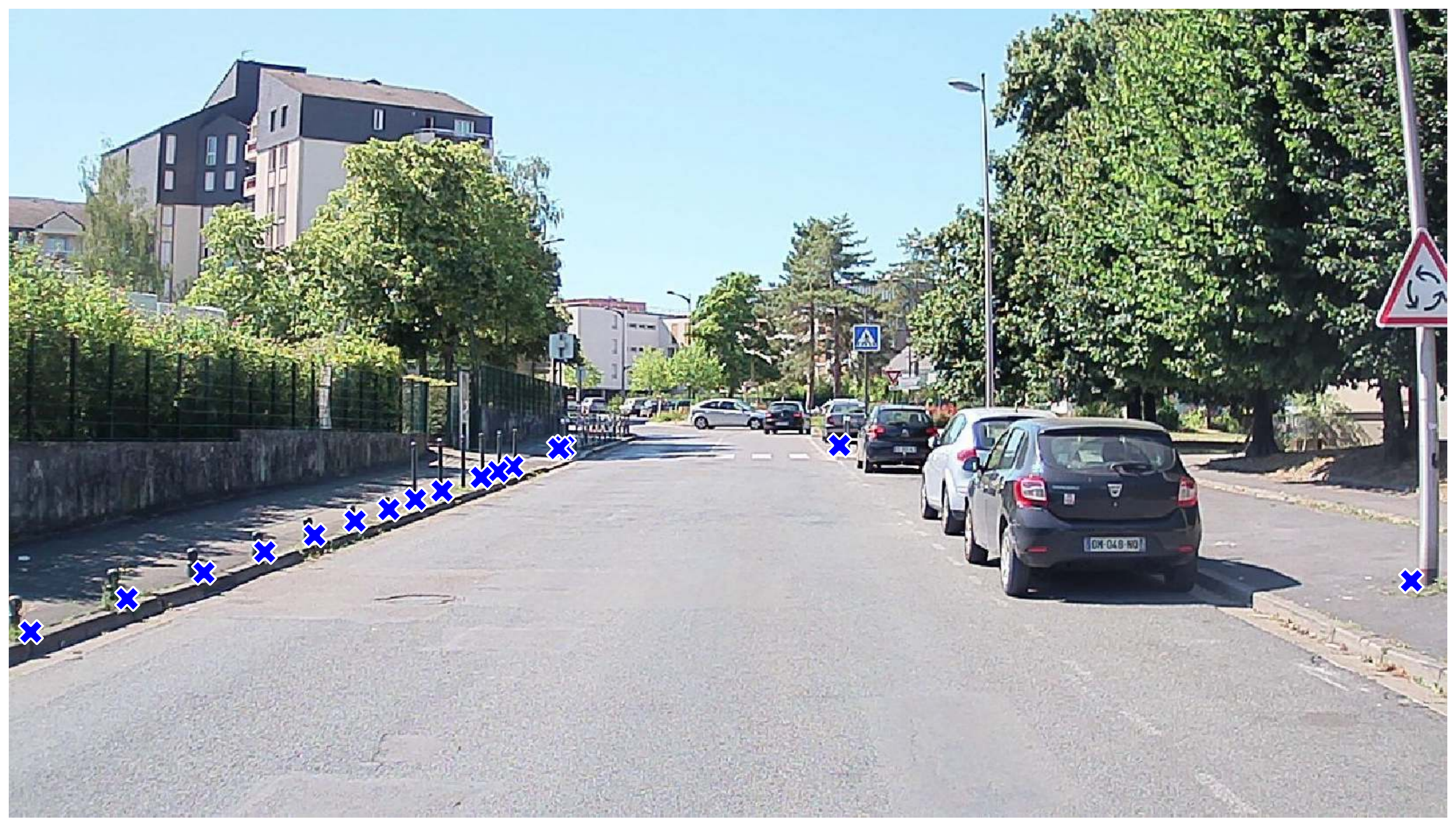}
 \end{subfigure}
 \begin{subfigure}[b]{0.19\textwidth}
  \includegraphics[width=\columnwidth]{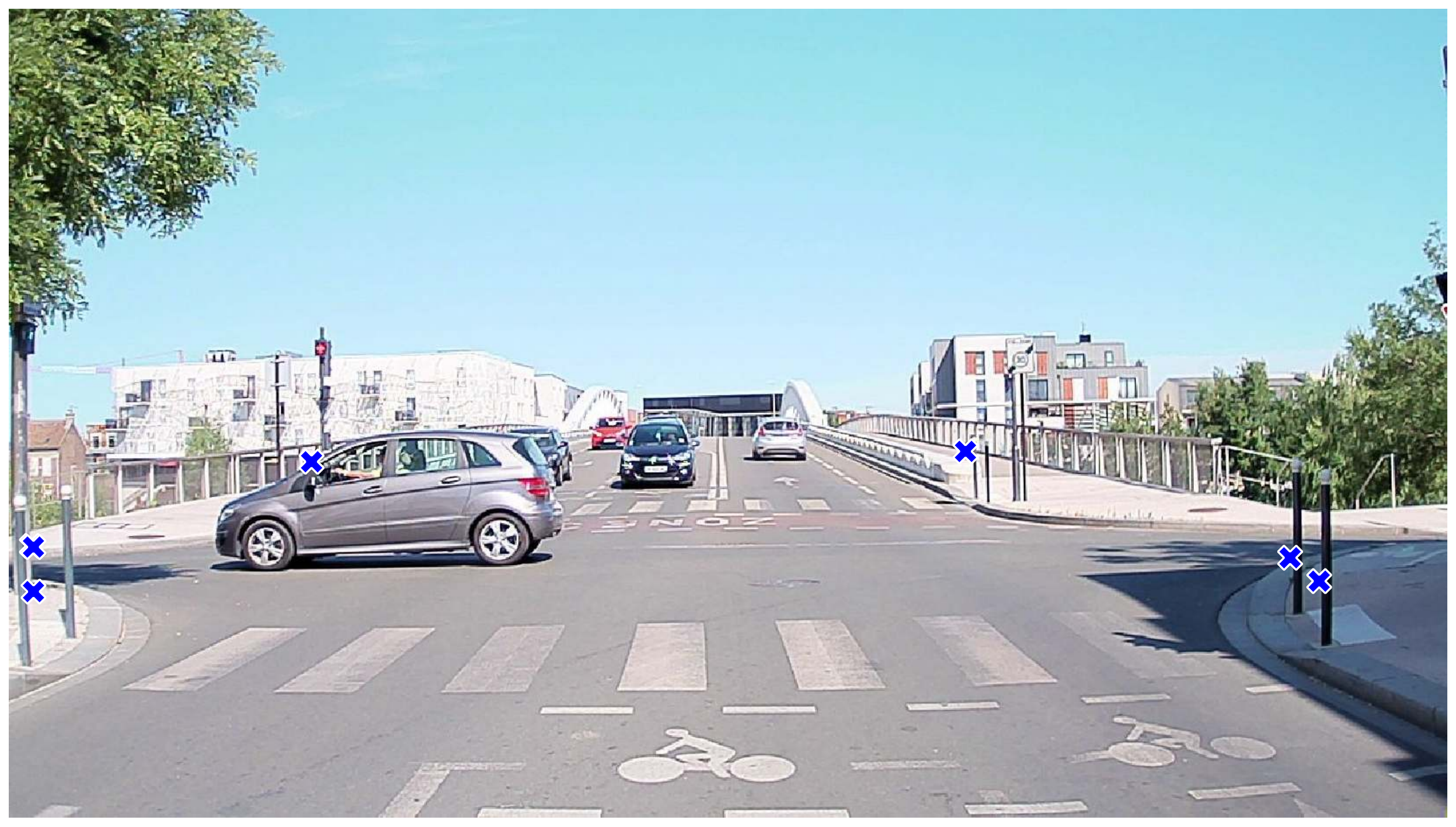}
 \end{subfigure}
 \begin{subfigure}[b]{0.19\textwidth}
  \includegraphics[width=\columnwidth]{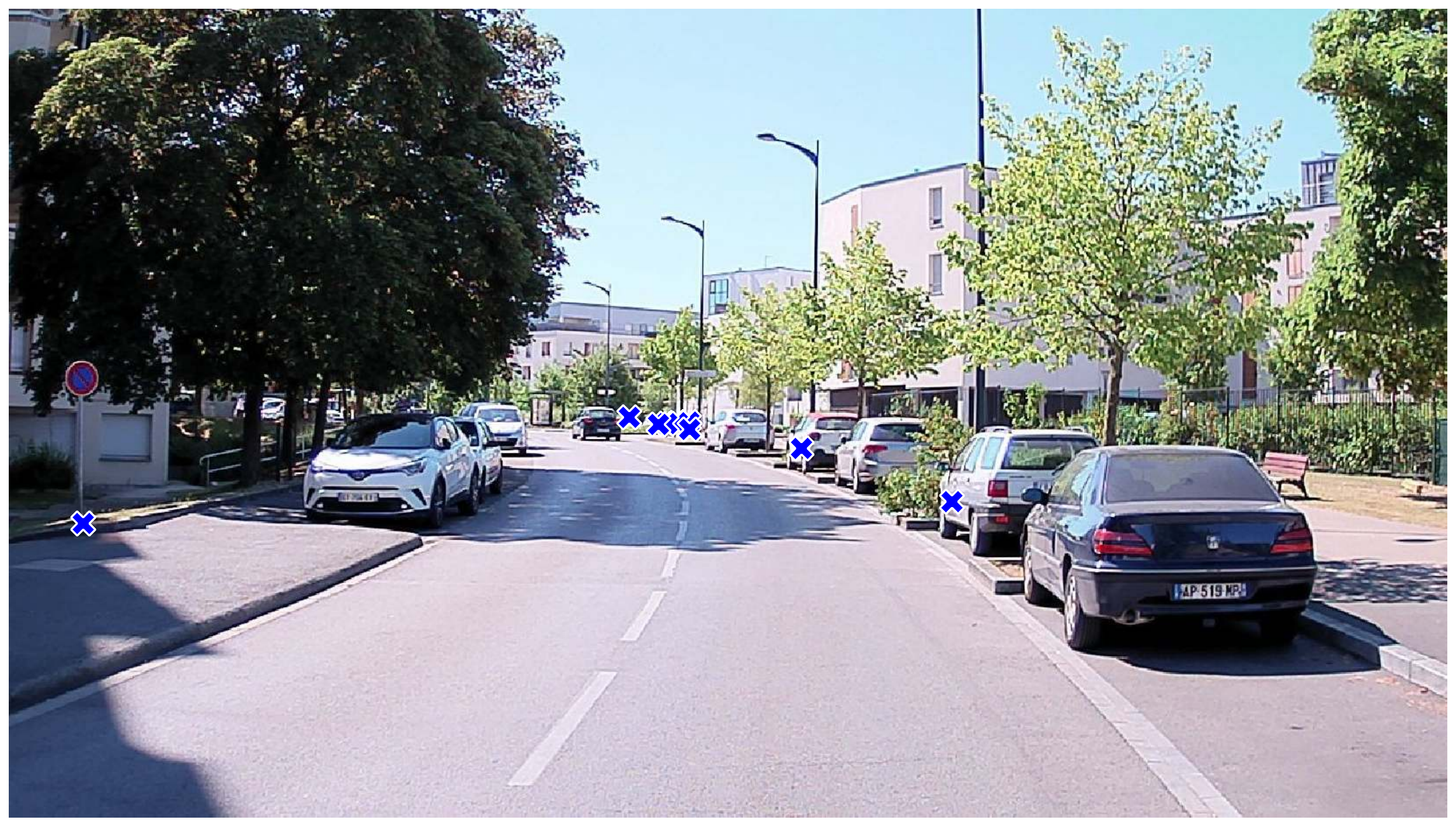}
 \end{subfigure}
 \begin{subfigure}[b]{0.19\textwidth}
  \includegraphics[width=\columnwidth]{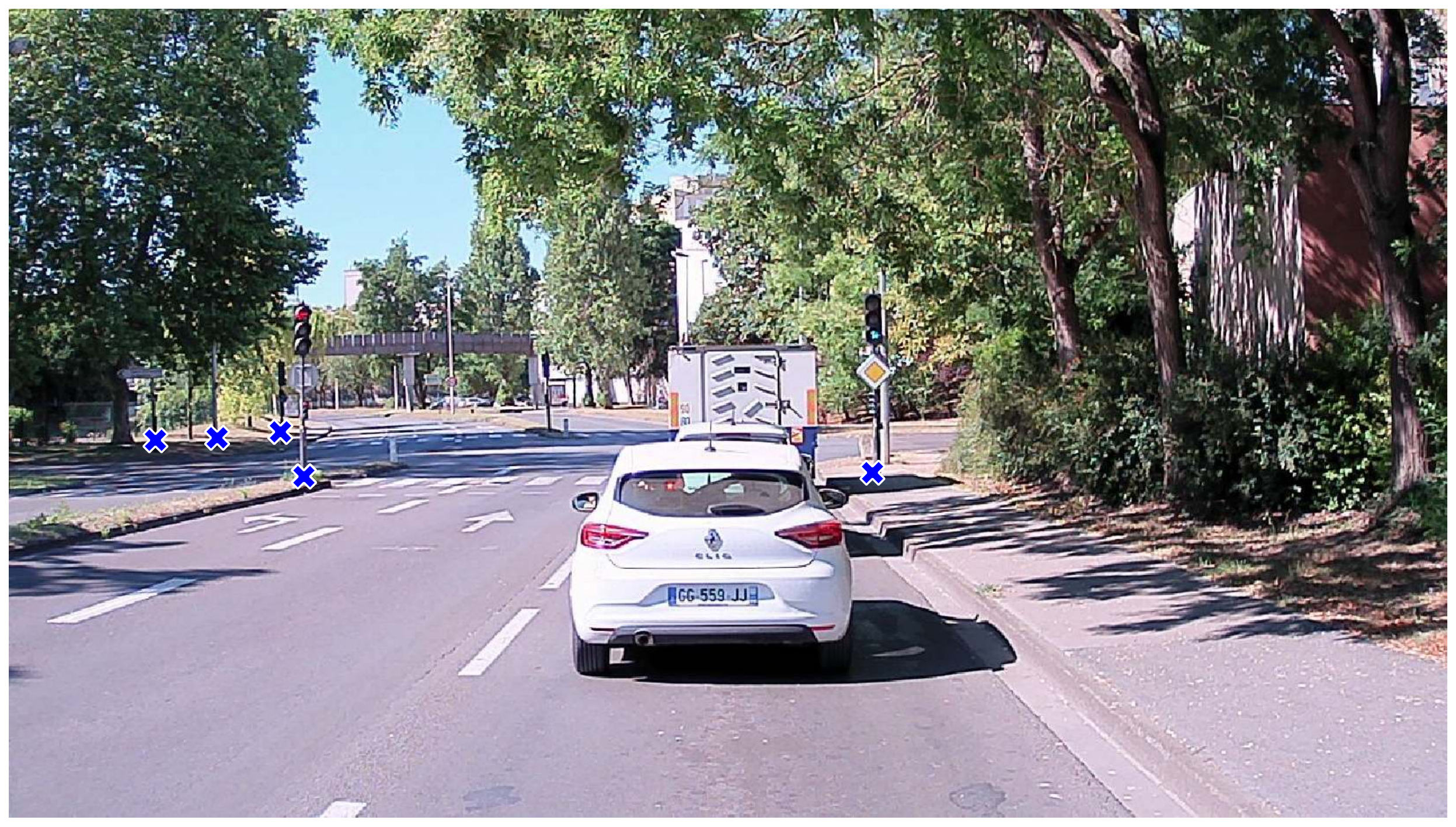}
 \end{subfigure}
 \begin{subfigure}[b]{0.19\textwidth}
  \includegraphics[width=\columnwidth]{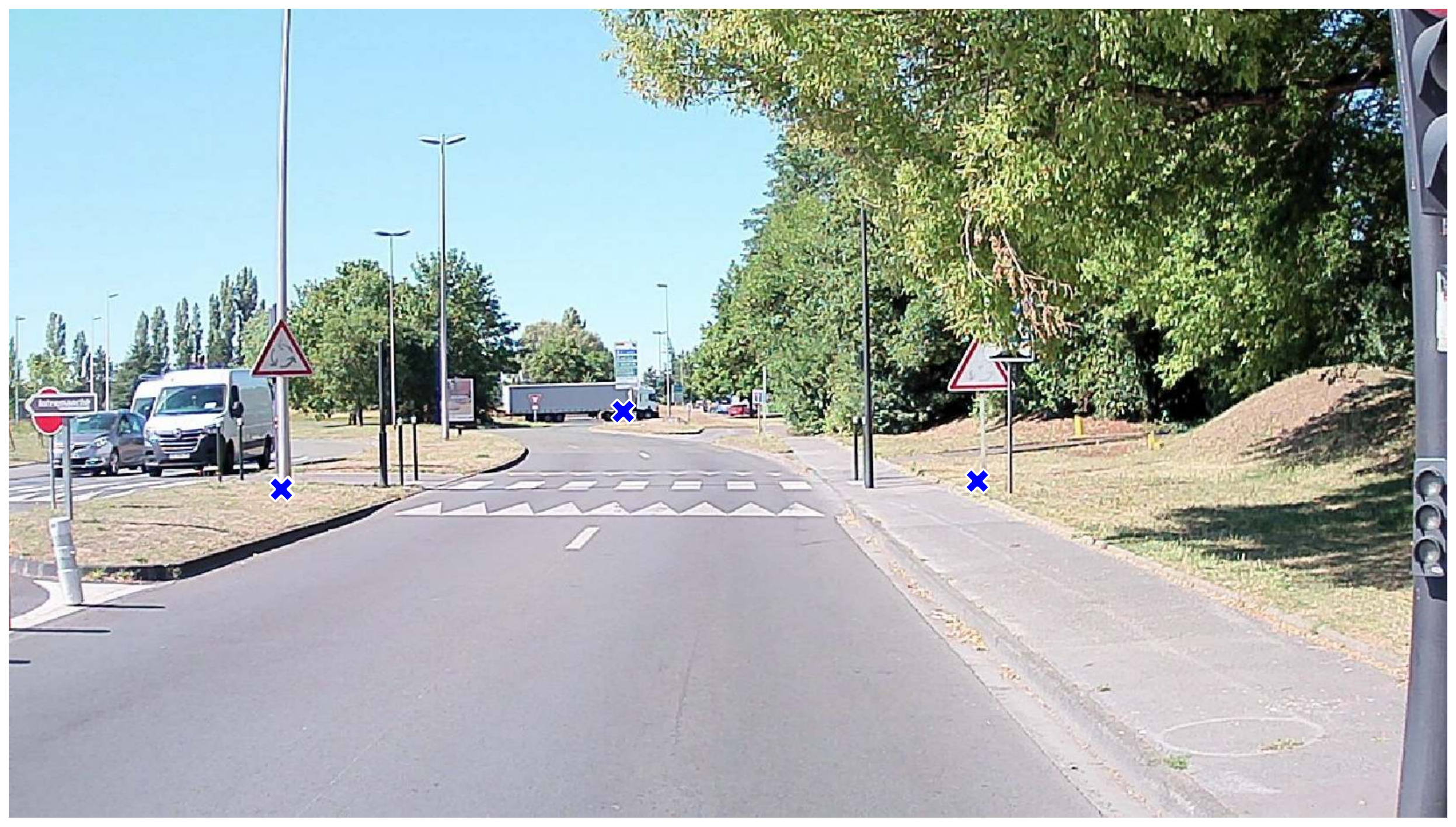}
 \end{subfigure}
 
 \begin{subfigure}[b]{0.19\textwidth}
  \includegraphics[width=\columnwidth]{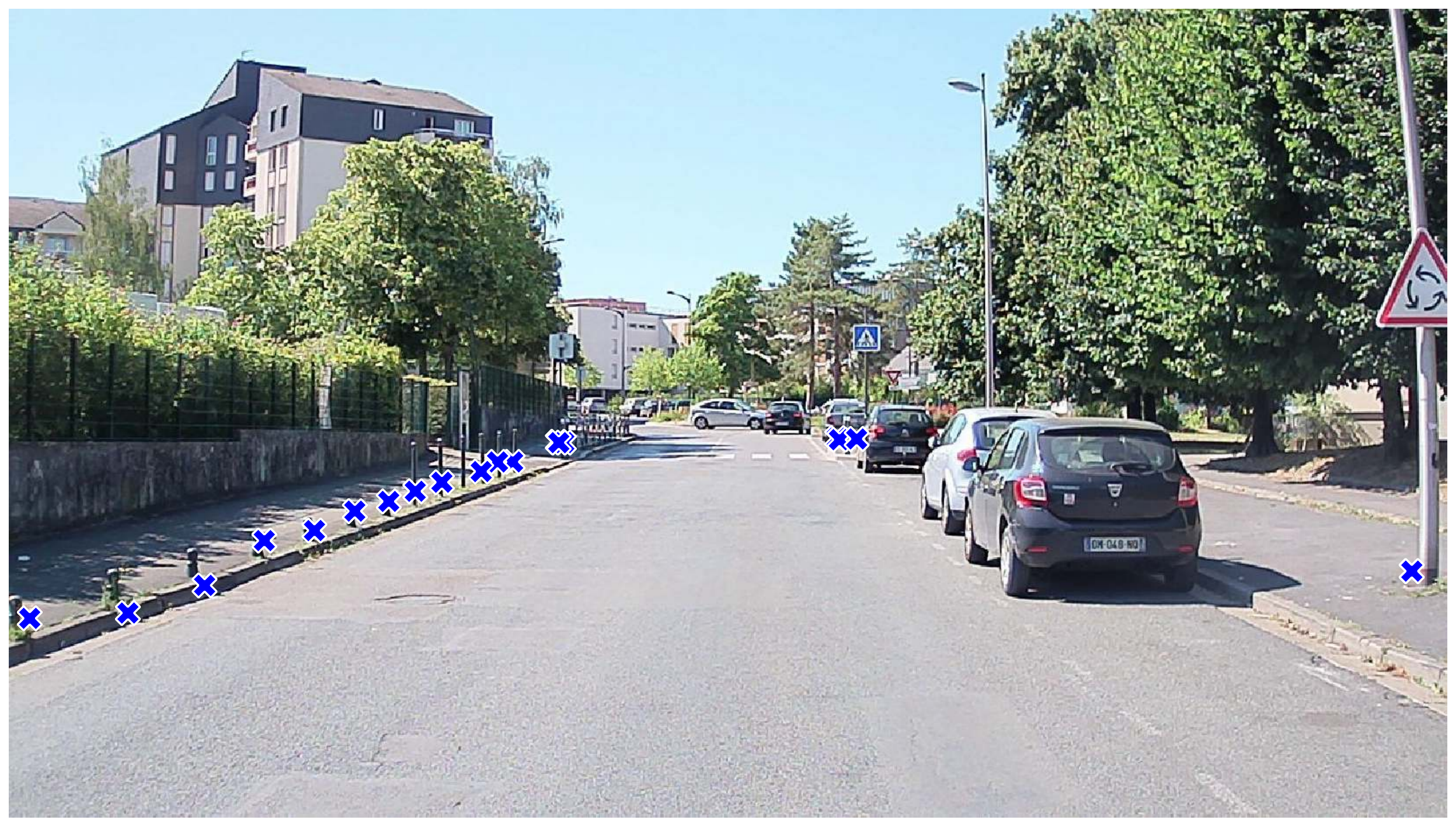}
  \caption{Nominal case}
 \end{subfigure}
 \begin{subfigure}[b]{0.19\textwidth}
  \includegraphics[width=\columnwidth]{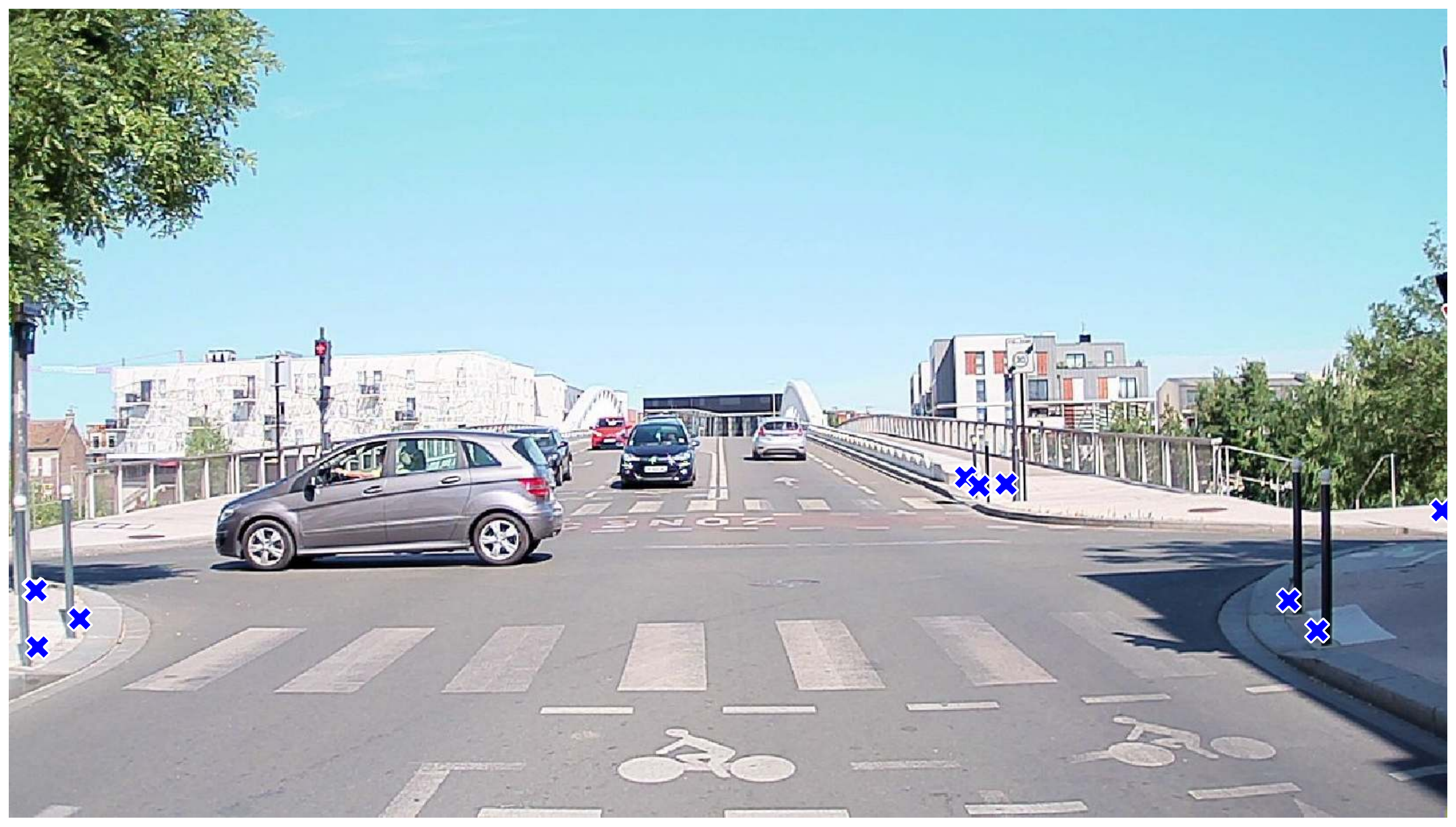}
  \caption{Good ground fitting}
 \end{subfigure}
 \begin{subfigure}[b]{0.19\textwidth}
  \includegraphics[width=\columnwidth]{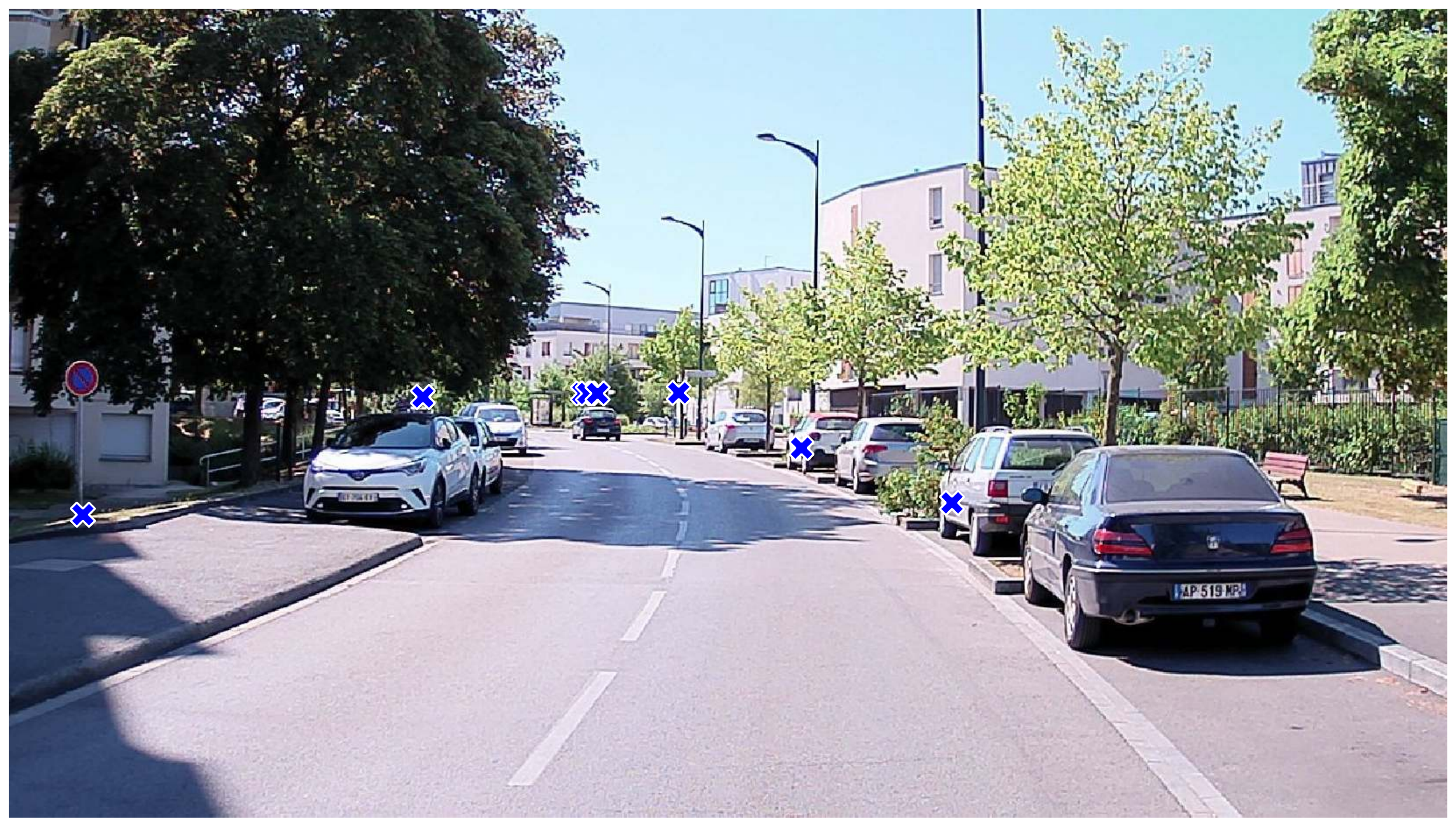}
  \caption{Unfiltered poles}
 \end{subfigure}
 \begin{subfigure}[b]{0.19\textwidth}
  \includegraphics[width=\columnwidth]{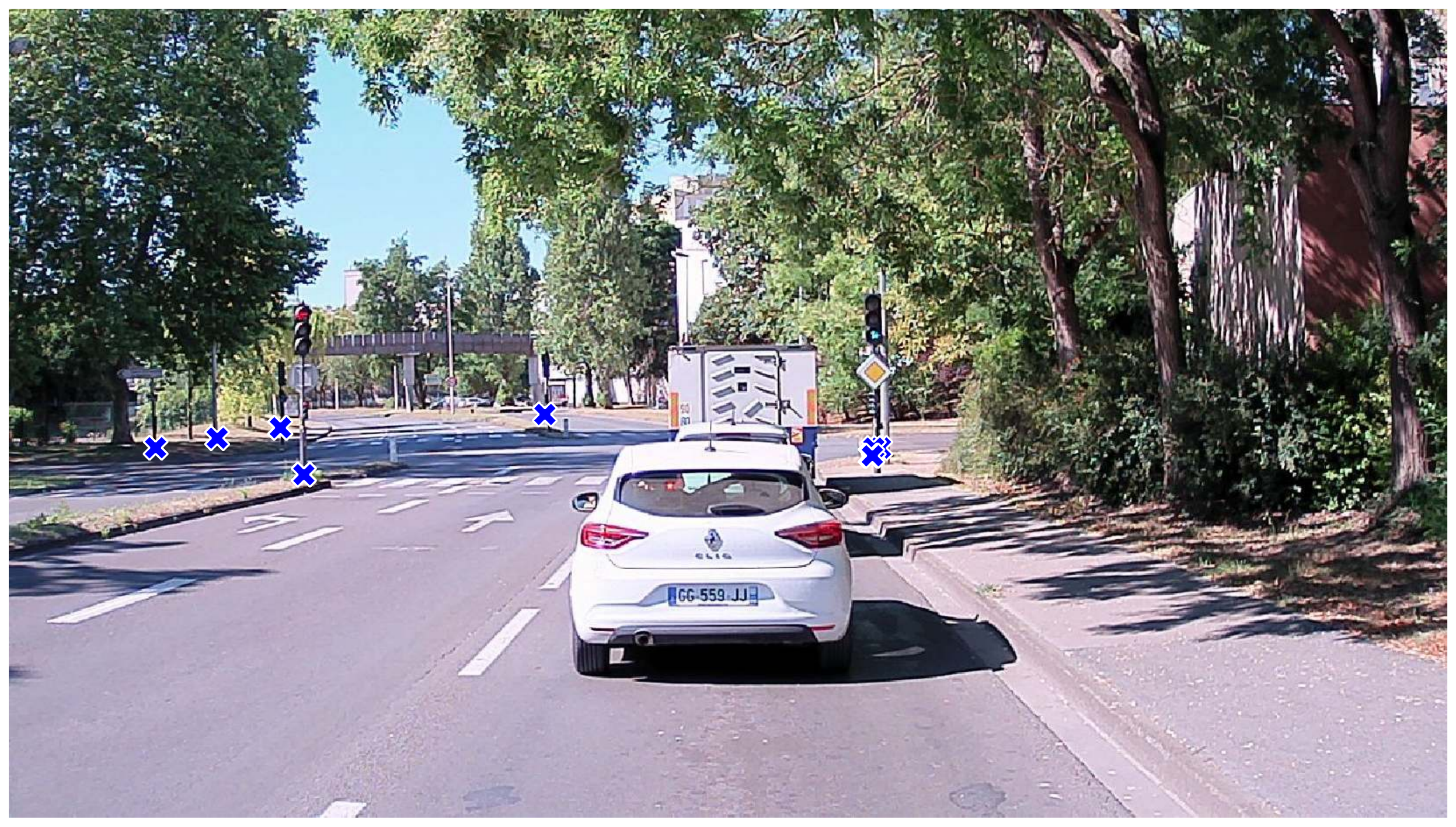}
  \caption{Non-annotated poles}
 \end{subfigure}
 \begin{subfigure}[b]{0.19\textwidth}
  \includegraphics[width=\columnwidth]{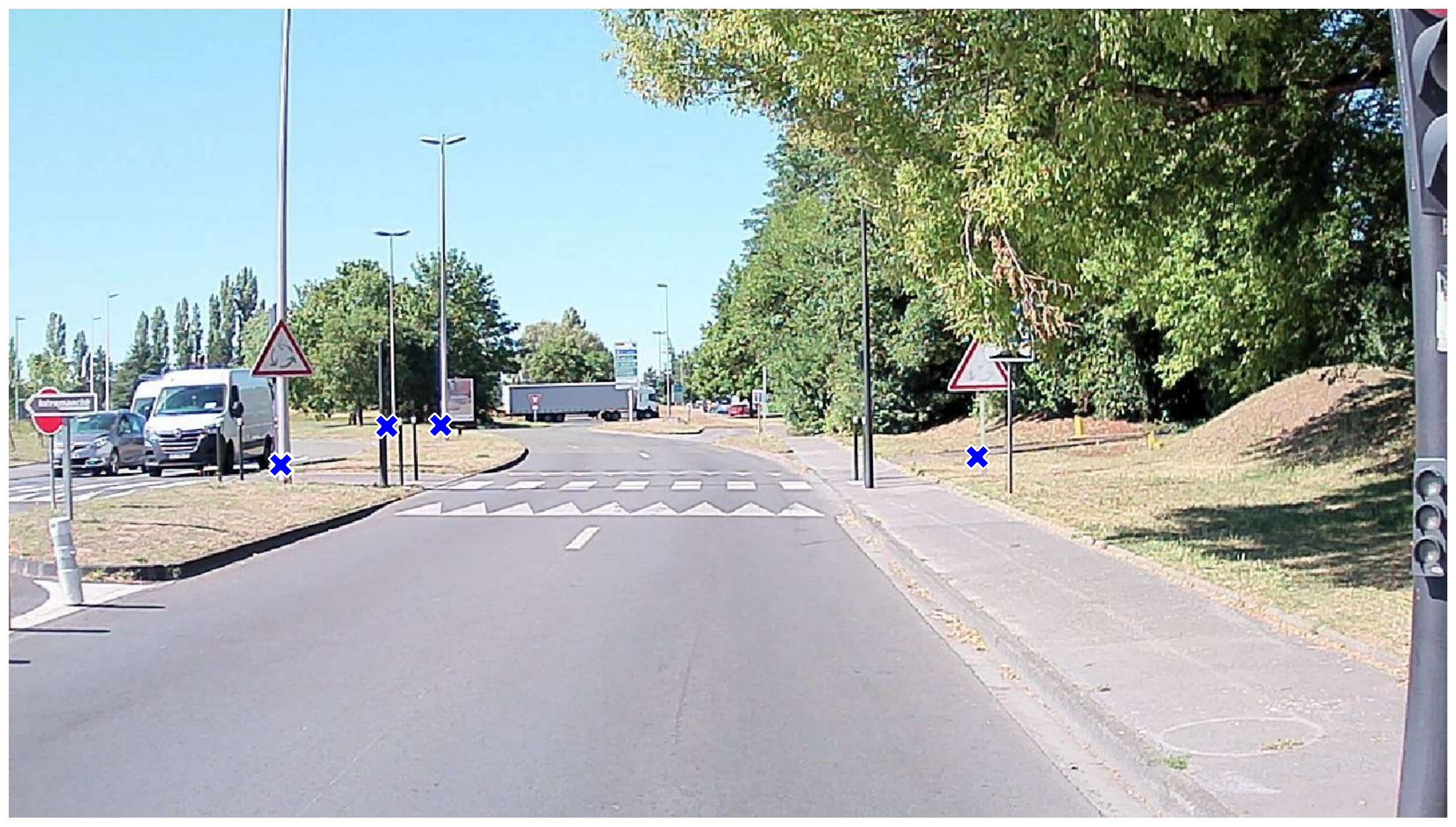}
  \caption{Unmapped features}
  \label{fig:unmapped}
 \end{subfigure}
 \caption{Examples of map-aided annotated images from our dataset in different situations. The first row corresponds to a naive projection of the HD map features with a simple filtering of far away features. The second row adds a lidar-based filtering to remove the occluded features. The last row adds an additional ground segmentation to refine the height estimation.}
 \label{fig:hds_annotation}
\end{figure*}

Fig.~\ref{fig:hds_annotation} illustrates examples of map-aided annotated images from our dataset in different situations. 
The first row corresponds to a naive projection of the HD map features with a simple filtering of far away features. 
The second row adds a lidar-based filtering to remove the occluded features. 
The last row adds an additional ground segmentation to refine the height estimation.
In average, adding each of these additional filtering improve the annotations but occasionally some noises can also be introduced due to wrongly segmented ground or bad depth estimation.
In addition, the unmapped features shown in Fig.~\ref{fig:unmapped} can not be dealt with our current approach.

Using these automatically annotated training data, we trained a model in the same way as we did on the BDD100K dataset.
Table~\ref{tab:hds-detect-metrics2} shows the detection performances on the Compi\`egne validation set comparing the model trained on BDD100K to the one trained in Compi\`egne using a 200x200 box size and the same confidence threshold 0.25 used previously.

\begin{table}[t!]
\caption{\label{tab:hds-detect-metrics2}Detection metrics after 100 epochs of training on the validation set of Compi\`egne using a 200x200 box size. Precision and recall are given for a confidence threshold of 0.25.} 
\smallskip
\begin{tabularx}{0.48\textwidth}{l*{4}{Y}}
 \toprule     
 Model &    mAP 0.5:0.95 &   Prec.  & Rec.  \\
 \cmidrule(lr){2-4}
 BDD100K    & 34.9\%           & 85.6\%         & 28.9\% \\
 Compi\`egne   & 31.6\%           & 63.9\%          & 50.1\%\\
\bottomrule
\end{tabularx}
\label{tab:hds}
\end{table}

\begin{figure}[t!]
 \centering
 \includegraphics[width=\columnwidth]{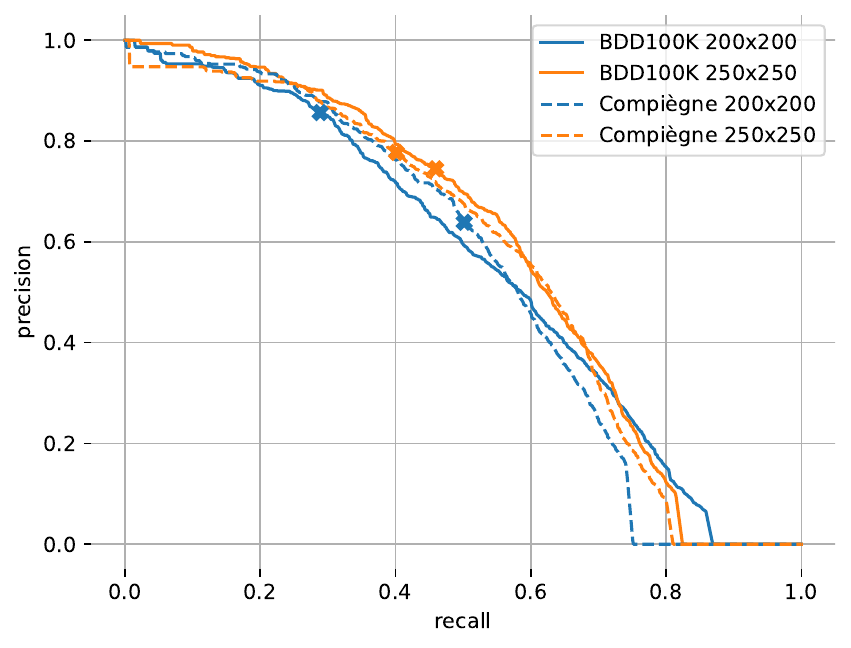}
 \caption{Precision-Recall curves obtained after 100 epochs of training with BDD100K and Compi\`egne models with 200x200 and 250x250 box sizes. The points corresponding to precision-recall values obtained using a confidence threshold of 0.25 are visible using crosses.}
 \label{fig:precision-recall-hds}
\end{figure}

The precision-recall curves for BDD100K and Compi\`egne models, utilizing 200x200 and 250x250 box sizes, are depicted in Figure~\ref{fig:precision-recall-hds}. As suggested by Table~\ref{tab:hds-detect-metrics2}, the BDD100K models exhibit distinct performance on the Compi\`egne validation set. For any precision objective above 0.25, the 250x250 model outperforms the other, illustrating the dependence of the box size choice on the studied validation set. Even for the Compi\`egne models, the 250x250 box size appears to be a preferable choice across various precision-recall trade-offs.


\begin{figure*}[t!]
 \centering
 \begin{subfigure}[b]{0.32\textwidth}
  \includegraphics[width=\columnwidth]{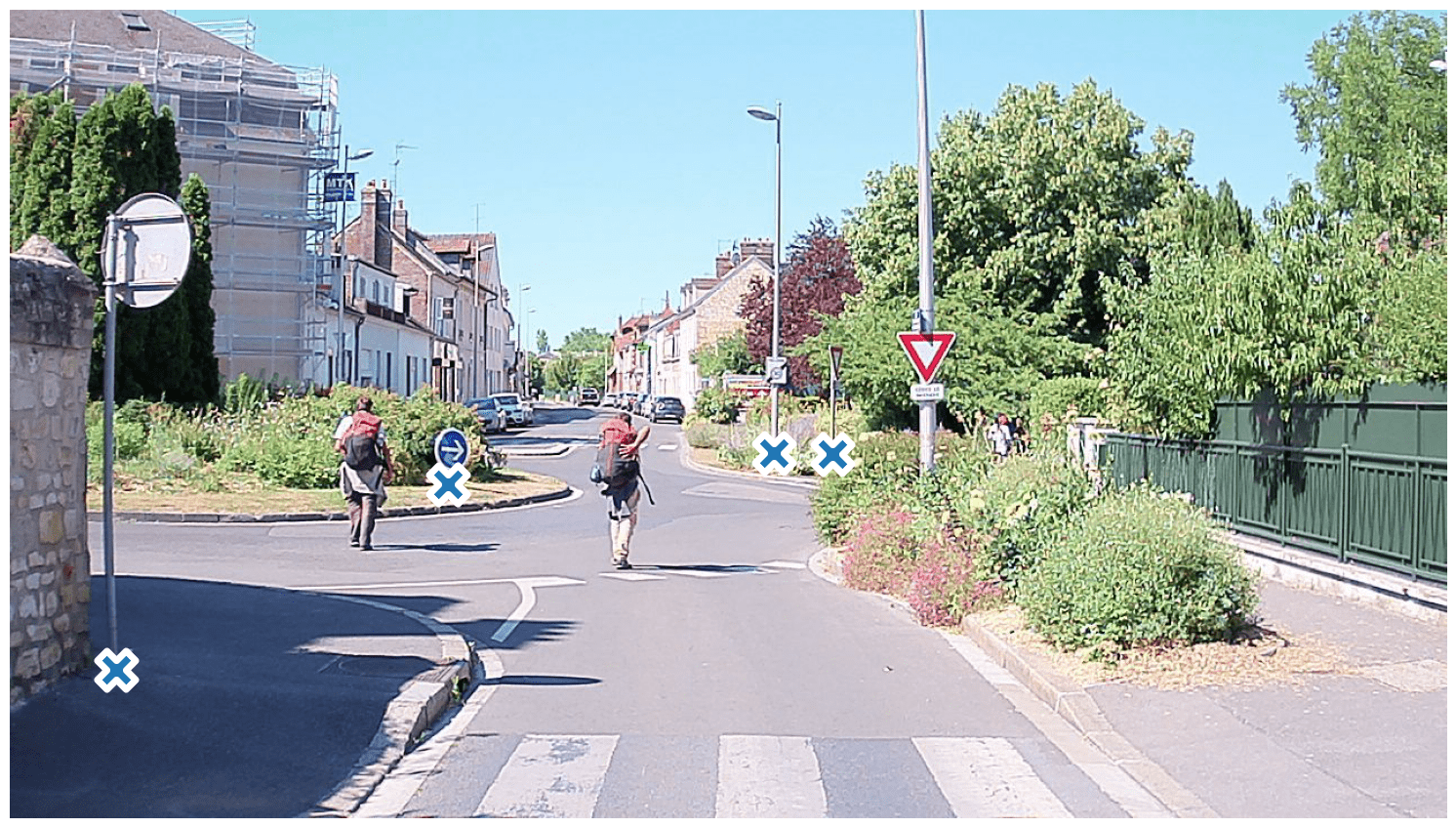}
  \caption{Ground truth}
 \end{subfigure}
 \begin{subfigure}[b]{0.32\textwidth}
  \includegraphics[width=\columnwidth]{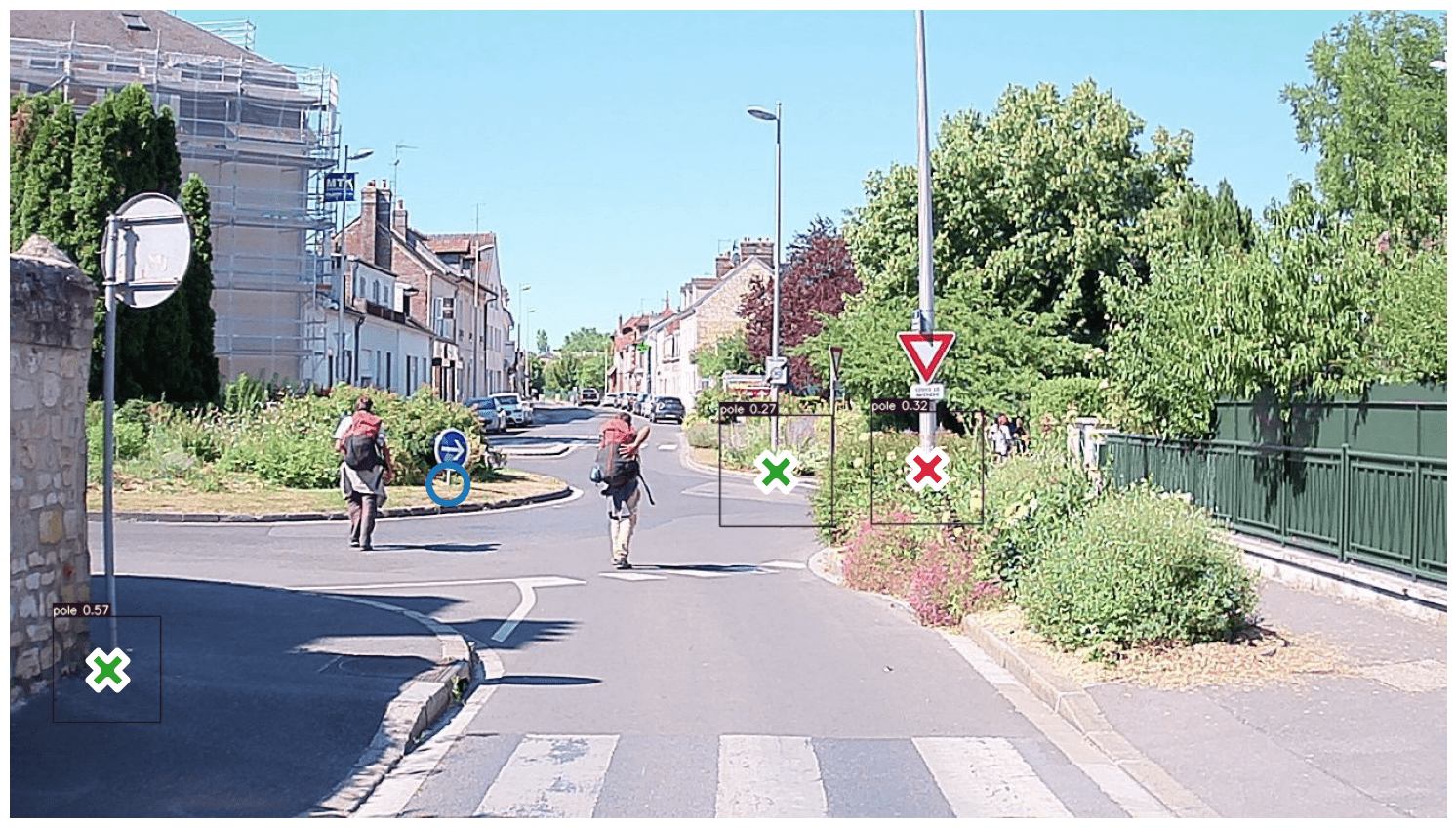}
  \caption{Prediction from BDD100K model}
 \end{subfigure}
 \begin{subfigure}[b]{0.32\textwidth}
  \includegraphics[width=\columnwidth]{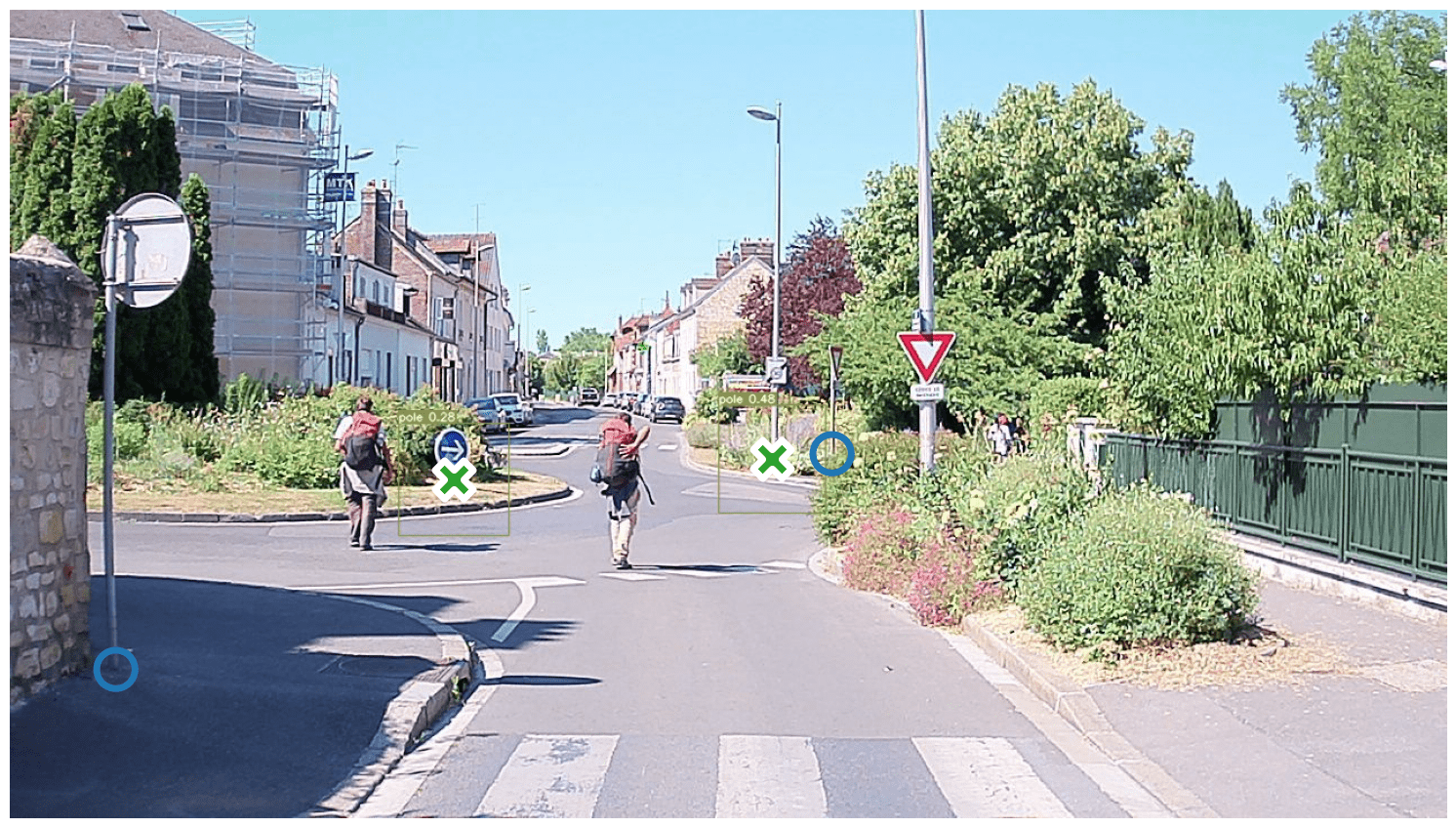}
  \caption{Prediction from Compi\`egne model}
 \end{subfigure}
 \caption{Ground truth and predictions from models trained on BDD100K and Compi\`egne on a Compi\`egne validation image.
 The green crosses are true positives, red ones false negatives and the blue circles are misdetections.}
 \label{fig:pred-compiegne}
\end{figure*}

Fig.~\ref{fig:pred-compiegne} illustrates a detection example using both 250x250 models on a Compi\`egne validation image with a confidence threshold set to 0.25.

We observe notable differences in the performance of the BDD100K model compared to the results presented in the previous section, characterized by high precision and poor recall. The lower mAP attained by this model suggests a distinct precision-recall curve compared to the previous observations. The optimal box size for this arbitrary confidence threshold may vary from the one used on the BDD100K validation set and even across different precision-recall trade-offs.
In contrast, the Compi\`egne model demonstrates satisfactory performance at this arbitrary threshold, showcasing the feasibility of training a pole base detector using automatic annotations generated from a map. However, the relatively low mAP indicates that performance may fluctuate when adjusting the confidence threshold or IoU threshold.

The main reason is likely to come from the imperfect annotations despite the lidar-based refinement and filtering.
A second reason may come from the low variability of the training data which was extracted from a single driving sequence.
However, the automatic annotation framework introduced in this work enables to easily add supplementary training data from new driving sequences without additional cost in terms of data labeling.

\section{Conclusion}
In this paper, we introduced a framework to use an HD map to automatically annotate images.
The map features were encoded as pole bases and projected onto images while a lidar was used to filter occluded features and to better estimate the ground surface.
Although a pole base was annotated by a single point, we demonstrated that it could be detected using a traditional object detector using bounding boxes.
This approach was first validated from annotations generated from semantic segmentation on the BDD100K.
The method was then tested using annotations generated from an HD map of the city of Compi\`egne.

This preliminary work opens room to further improvements.
The lidar-based filtering and refinement could be improved by tuning better the different parameters, but also with more sophisticated geometric consideration.
The training data could be subsampled in a same way as the validation data to avoid training on too similar images with potentially wrong automated annotations leading to a loss of performance.
The automated annotations, which are still imperfect, can also serve as an initial annotation either for a human operator or as an input for a second stage refinement.
The ability to have a low-cost annotation pipeline could be exploited in many contexts such as generating training data in different driving weather or lighting conditions without having the need to label the data from scratch.
The HD map used in this work also contained other features such as lane markings.
This approach could also be extended to annotate data to train a lane marking detector.
Finally, the use of the proposed pole base detector will be studied within a localization context in future work.

\section*{Acknowledgment}
This work has been funded by the European project ERASMO~\cite{ERASMO23} (GSA/GRANT/03/2018) in the framework of the SIVALab laboratory between Renault and Heudiasyc.
The authors would like to thank R\'emy Huet and Vincent Brebion for their technical support in this work.
\bibliographystyle{IEEEtran}
\bibliography{IEEEabrv,biblio}

\begin{thebibliography}{10}
\providecommand{\url}[1]{#1}
\csname url@rmstyle\endcsname
\providecommand{\newblock}{\relax}
\providecommand{\bibinfo}[2]{#2}
\providecommand\BIBentrySTDinterwordspacing{\spaceskip=0pt\relax}
\providecommand\BIBentryALTinterwordstretchfactor{4}
\providecommand\BIBentryALTinterwordspacing{\spaceskip=\fontdimen2\font plus
\BIBentryALTinterwordstretchfactor\fontdimen3\font minus
  \fontdimen4\font\relax}
\providecommand\BIBforeignlanguage[2]{{%
\expandafter\ifx\csname l@#1\endcsname\relax
\typeout{** WARNING: IEEEtran.bst: No hyphenation pattern has been}%
\typeout{** loaded for the language `#1'. Using the pattern for}%
\typeout{** the default language instead.}%
\else
\language=\csname l@#1\endcsname
\fi
#2}}

\bibitem{IV23}
B.~Missaoui, M.~Noizet, and P.~Xu, ``Map-aided annotation for pole base
  detection,'' in \emph{IEEE Intelligent Vehicles Symposium Workshop}, June
  2023.

\bibitem{li_robust_2021}
L.~Li, M.~Yang, L.~Weng, and C.~Wang, ``\BIBforeignlanguage{en}{Robust
  localization for intelligent vehicles based on pole-like features using the
  point cloud},'' \emph{\BIBforeignlanguage{en}{IEEE Transactions on Automation
  Science and Engineering}}, pp. 1--14, 2021.

\bibitem{sefati_improving_2017}
M.~Sefati, M.~Daum, B.~Sondermann, K.~D. Kreiskother, and A.~Kampker,
  ``\BIBforeignlanguage{en}{Improving vehicle localization using semantic and
  pole-like landmarks},'' in \emph{\BIBforeignlanguage{en}{{IEEE} {Intelligent}
  Vehicles Symposium}}, Los Angeles, CA, USA, June 2017, pp. 13--19.

\bibitem{spangenberg_pole-based_2016}
R.~Spangenberg, D.~Goehring, and R.~Rojas, ``\BIBforeignlanguage{en}{Pole-based
  localization for autonomous vehicles in urban scenarios},'' in
  \emph{\BIBforeignlanguage{en}{{IEEE}/{RSJ} {International} {Conference} on
  {Intelligent} {Robots} and {Systems}}}, Daejeon, South Korea, Oct. 2016, pp.
  2161--2166.

\bibitem{gouda_fully_2022}
M.~Gouda, A.~Shalkamy, X.~Li, and K.~El-Basyouny,
  ``\BIBforeignlanguage{en}{Fully automated algorithm for light pole detection
  and mapping in rural highway environment using mobile light detection and
  ranging point clouds},'' \emph{\BIBforeignlanguage{en}{Transportation
  Research Record: Journal of the Transportation Research Board}}, vol. 2676,
  no.~7, pp. 617--629, July 2022.

\bibitem{lehtomaki_detection_2010}
M.~Lehtomäki, A.~Jaakkola, J.~Hyyppä, A.~Kukko, and H.~Kaartinen,
  ``\BIBforeignlanguage{en}{Detection of vertical pole-like objects in a road
  environment using vehicle-based laser scanning data},''
  \emph{\BIBforeignlanguage{en}{Remote Sensing}}, vol.~2, no.~3, pp. 641--664,
  Feb. 2010.

\bibitem{rodriguez-cuenca_automatic_2015}
B.~Rodríguez-Cuenca, S.~García-Cortés, C.~Ordóñez, and M.~Alonso,
  ``\BIBforeignlanguage{en}{Automatic detection and classification of pole-like
  objects in urban point cloud data using an anomaly detection algorithm},''
  \emph{\BIBforeignlanguage{en}{Remote Sensing}}, vol.~7, no.~10, pp.
  12\,680--12\,703, Sept. 2015.

\bibitem{ghallabi_lidar-based_2019}
F.~Ghallabi, G.~El-Haj-Shhade, M.-A. Mittet, and F.~Nashashibi,
  ``\BIBforeignlanguage{en}{Lidar-based road signs detection for vehicle
  localization in an {HD} map},'' in \emph{\BIBforeignlanguage{en}{{IEEE}
  {Intelligent} {Vehicles} {Symposium}}}, Paris, France, June 2019, pp.
  1484--1490.

\bibitem{lin14}
T.-Y. Lin, M.~Maire, S.~Belongie, J.~Hays, P.~Perona, D.~Ramanan,
  P.~Doll{\'a}r, and C.~L. Zitnick, ``Microsoft {COCO}: Common objects in
  context,'' in \emph{Computer Vision -- ECCV}, D.~Fleet, T.~Pajdla,
  B.~Schiele, and T.~Tuytelaars, Eds.\hskip 1em plus 0.5em minus 0.4em\relax
  Cham: Springer International Publishing, 2014, pp. 740--755.

\bibitem{geiger12}
A.~Geiger, P.~Lenz, and R.~Urtasun, ``Are we ready for autonomous driving?
  {T}he {KITTI} vision benchmark suite,'' in \emph{IEEE Conference on Computer
  Vision and Pattern Recognition}, 2012, pp. 3354--3361.

\bibitem{Behley19}
J.~Behley, M.~Garbade, A.~Milioto, J.~Quenzel, S.~Behnke, C.~Stachniss, and
  J.~Gall, ``Semantic{KITTI}: A dataset for semantic scene understanding of
  lidar sequences,'' in \emph{Proceedings of the IEEE/CVF International
  Conference on Computer Vision}, October 2019.

\bibitem{cordts16}
M.~Cordts, M.~Omran, S.~Ramos, T.~Rehfeld, M.~Enzweiler, R.~Benenson,
  U.~Franke, S.~Roth, and B.~Schiele, ``The {Cityscapes} dataset for semantic
  urban scene understanding,'' in \emph{Proceedings of the IEEE Conference on
  Computer Vision and Pattern Recognition}, June 2016, pp. 3212--3223.

\bibitem{yu20}
F.~Yu, H.~Chen, X.~Wang, W.~Xian, Y.~Chen, F.~Liu, V.~Madhavan, and T.~Darrell,
  ``{BDD100K}: A diverse driving dataset for heterogeneous multitask
  learning,'' in \emph{Proceedings of the IEEE/CVF Conference on Computer
  Vision and Pattern Recognition}, June 2020, pp. 2636--2645.

\bibitem{dong23}
H.~Dong, X.~Chen, S.~Särkkä, and C.~Stachniss, ``Online pole segmentation on
  range images for long-term lidar localization in urban environments,''
  \emph{Robotics and Autonomous Systems}, vol. 159, p. 104283, 2023.

\bibitem{Sun20}
C.~Sun, J.~M.~U. Vianney, Y.~Li, L.~Chen, L.~Li, F.-Y. Wang, A.~Khajepour, and
  D.~Cao, ``Proximity based automatic data annotation for autonomous driving,''
  \emph{IEEE/CAA Journal of Automatica Sinica}, vol.~7, no.~2, pp. 395--404,
  2020.

\bibitem{Lee21}
W.~H. Lee, K.~Jung, C.~Kang, and H.~S. Chang, ``Semi-automatic framework for
  traffic landmark annotation,'' \emph{IEEE Open Journal of Intelligent
  Transportation Systems}, vol.~2, pp. 1--12, 2021.

\bibitem{lee22}
S.~Lee, H.~Lim, and H.~Myung, ``Patchwork++: Fast and robust ground
  segmentation solving partial under-segmentation using 3d point cloud,'' in
  \emph{IEEE/RSJ International Conference on Intelligent Robots and Systems},
  2022, pp. 13\,276--13\,283.

\bibitem{wang22}
C.-Y. Wang, A.~Bochkovskiy, and H.-Y.~M. Liao, ``{YOLO}v7: Trainable
  bag-of-freebies sets new state-of-the-art for real-time object detectors,''
  \emph{arXiv preprint:2207.02696}, 2022.

\bibitem{ERASMO23}
L.~Vilalta~Estrada, C.~Mu\~noz Garc\'ia, E.~Dom\'inguez~Tijero, M.~Noizet,
  P.~Xu, S.~Y. Voon, S.~Guerassimov, and W.~W.~Cox, ``{ERASMO} -- {E}nhanced
  {R}eceiver for {A}utonomou{S} {MO}bility,'' in \emph{Proceedings of the 15th
  ITS European Congress}, May 2023.

\end{thebibliography}
\end{document}